\documentclass[%
 reprint,
superscriptaddress,
 amsmath,amssymb,
 aps,
floatfix,
]{revtex4-2}

\usepackage{graphicx}% Include figure files
\usepackage{dcolumn}% Align table columns on decimal point
\usepackage{bm}% bold math
\usepackage{siunitx}
\usepackage{float}
\usepackage{natbib}
\usepackage[pagewise]{lineno}
\begin{document}

\preprint{APS/123-QED}

\title{Geometry-Induced Dynamics of Confined Chiral Active Matter} % Force line breaks with \\

\author{Archit Negi}
\affiliation{%
 Department of Physics, Kyushu University, Motooka 744, Fukuoka 819-0395, Japan
}%

% \collaboration{MUSO Collaboration}%\noaffiliation

\author{Kazusa Beppu}
\affiliation{%
 Department of Physics, Kyushu University, Motooka 744, Fukuoka 819-0395, Japan
}
\affiliation{Department of Applied Physics, Aalto University School of Science, Puumiehenkuja 2, Espoo, 02150, Finland}

%
% \affiliation{
%  Third institution, the second for Charlie Author
% }%
\author{Yusuke T. Maeda}
 \email{ymaeda@phys.kyushu-u.ac.jp}
 \affiliation{%
 Department of Physics, Kyushu University, Motooka 744, Fukuoka 819-0395, Japan
}%

% \collaboration{CLEO Collaboration}%\noaffiliation

\date{\today}% It is always \today, today,

\begin{abstract}
Controlling the motion of active matter is a central issue that has recently garnered significant attention in fields ranging from non-equilibrium physics to chemical engineering and biology. Distinct methods for controlling active matter have been developed, and physical confinement to limited space and active matter with broken rotational symmetry (chirality) are two prominent mechanisms. However, the interplay between pattern formation due to physical constraints and the ordering by chiral motion needs to be better understood. In this study, we conduct numerical simulations of chiral self-propelled particles under circular boundary confinement. The collective motion of confined self-propelled particles can take drastically different forms depending on their chirality. The balance of orientation changes between particle interaction and the boundary wall is essential for generating ordered collective motion. Our results clarify the role of the steric boundary effect in controlling chiral active matter.
% \begin{description}
% Secondary publications and information retrieval purposes.
% You may use the \texttt{description} environment to structure your abstract;
% use the optional argument of the \verb+\item+ command to give the category of each item. 
% \end{description}
\end{abstract}

\maketitle

%\tableofcontents

\section{\label{sec:level1}Introduction}
Active matter encompasses a broad range of systems with many constituent elements that consume energy for motion or exerting forces \cite{ramaswamy2010, marchetti2013, shankar2022}. These are inherently far from equilibrium systems which, thanks to the interactions between the individual components, can show collective motion and are found across the spatiotemporal scale; molecular motor proteins \cite{leibler1997, bausch2010, dogic2012}, bacterial turbulence \cite{aronson2022}, epithelial cell migration \cite{amin2018}, schools of fish \cite{guttal2020}, flocks of birds \cite{giardina2014} and crowds of people \cite{bartolo2019} are typical examples of active matter. In addition to being a highly fascinating field in and of itself, active matter has numerous potential applications as well \cite{leonardo2017, thomson2019}, mainly due to its ability to form self-sustained ordered structures and to fully realize that potential, control over its dynamics is essential.

Since local orientation interactions drive the collective motion of active matter, developing methods to manipulate the orientation of each particle is fundamental to tailoring their collectively ordered patterns. In particular, confinement of active matter has the ability to drastically alter its dynamics, such as the autonomous circulation of active fluids \cite{goldstein2012}, and over the years, physical geometric confinement has been shown to be a promising control mechanism for driving its organization. Numerical simulations of active fluids confined inside channels have been found to show channel width-dependent behavior, including boundary flows, vortex formation, and turbulent flow \cite{goldstein2013, goldstein2014, Shendruk2017, Doostmohammadi2017, Huang2021}. Confining active matter with polar orientation interactions to a circular space can, depending on the length scales of the confined elements and the confinement, transform active turbulent flows into an ordered global vortex state for bacterial suspensions \cite{goldstein2013, goldstein2016a, goldstein2016b, Beppu2017, nishiguchi2018, aronson2020, Beppu2021}, self-gliding microtubules \cite{Opathalage2019, Guillamat2017, araki2021}, and epithelial cells \cite{Doxzen2013}. The boundary shape aligns the orientation of a group of active matter, enabling control over the pattern of collective motion. Not only for active polar fluids, there is also the possibility of the formation of dynamic topological defects \cite{DeCamp2015, silberzan2018, roux2022} in dense active matter with nematic interaction, and if such systems are confined, the defects can drive the contractile or extensile flows \cite{Keber2014, Hsu2022}. However, even though the shape of individual bacteria or cells can affect collective dynamics, the motion of active matter in these models is often simplified as a particle moving straight ahead.

One such microscopic nature of how active matter affects collective motion is chirality \cite{libchen2022}. The individual elements in chiral active matter systems tend to move along a circular trajectory. Chirality \( (\omega) \) is thus the measure of the angular frequency of the circular motion. In a more general sense, there is a microscopic symmetry breaking of the handedness of these systems, which leads to a preferred direction of motion in their macroscopic dynamics \cite{caprini2019, caprini2021, snezhko2020}. Experimental realizations of chiral active matter can be in the form of particles having a chiral structure, such as chiral microswimmers \cite{Kmmel2013}, or pear-shaped colloidal rollers \cite{Zhang2020}. Self-propelled, self-spinning robots are also a kind of chiral active matter \cite{Scholz2018, Yang2020, Liu2020}, as are circle swimming bacteria \cite{Beppu2017, Beppu2021}. Microtubules can also be prepared in a way to show density-dependent chirality in \textit{in vitro} systems \cite{Afroze2021}. Studies about the chiral active matter, both numerical and experimental, show the variety in the dynamics of such systems, which is significantly different from achiral systems: in simulations of an unconfined, single-frequency chiral system, large rotating droplets or small flocks can form, depending on chirality and system density \cite{Liebchen2017}, while similar systems with multiple frequencies can show chirality dependent self-sorting and synchronization \cite{Levis2019a, Levis2019b}. As additional phases such as vortices \cite{Liao2021, Ventejou2021, Kruk2020} and bands \cite{Ventejou2021, Kruk2020} have also been found, there is a growing understanding of the new role of chirality in controlling orientation interactions of active systems.

However, most of the current studies on the chiral active matter focus on the dynamics in bulk, and the effects of confinement on the chiral collective behavior remain little understood. One study that considered self-propelled robots with and without chirality, confined to a circular area, showed that chirality suppresses cluster formation at the boundary \cite{Deblais2018}. Self-spinning confined rotors have been shown to exhibit boundary flows \cite{Yang2020, Liu2020}. Additionally, while the mixture of opposite rotating confined robots has been shown to phase separate \cite{Scholz2018}, the orientation interactions between particles with chirality and the changes in collective motion induced by their interactions with the wall have not been fully explored. To address this question in the present study, we investigate whether the chiral active matter can be further controlled through physical confinement and how the dynamics of such confined systems change with the different system parameters.

\section{Methods}
\begin{figure}[b]
\includegraphics[width=\columnwidth]{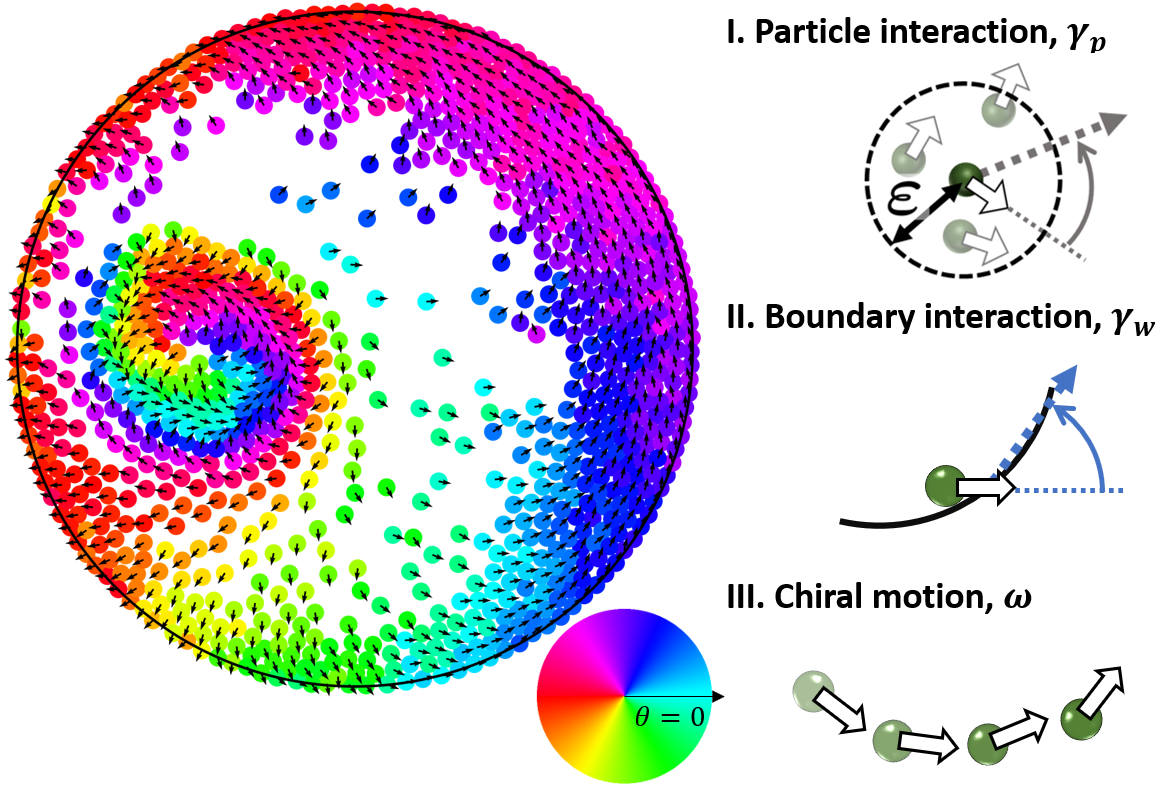}
\caption{
\label{fig:fig1}
\small{Schematic of the simulation system: chiral active particles, moving with a constant speed \(v_0\) and having an inherent chirality \(\omega\) are bound by a circular boundary. There is soft repulsion between the particles and between the particles and the wall. The particles have a polar interaction with each other (\(\gamma_p\)), and they try to align their orientation to the average orientation within the radius of interaction (\(\epsilon\)). Particles also interact nematically with the boundary (\(\gamma_w\)) and try to align to the tangential direction on collision with the boundary. Particle color denotes the orientation.}
}
\end{figure}

In this work, we do numerical simulations of chiral active matter confined to a limited circular space by extending a model of our previous study \cite{Beppu2021} (Fig.~\ref{fig:fig1}). Our system consists of \( N \) particles, each moving with a constant speed \( v_{0} \), inside a circular boundary of radius \( R \). The particles have a polar interaction with each other, meaning every particle wants its orientation to be equal to the mean orientation of its nearest neighbours; the strength of the polar interaction is given by \( \gamma_{p} \) (Fig.~\ref{fig:fig1}(I)). However, the random noise in the system, given by \( \eta (t) \), prevents perfect alignment between particles. The particles also interact nematically with the boundary, aligning parallel to it after a collision; the strength of the nematic interaction is given by \( \gamma_{w} \) (Fig.~\ref{fig:fig1}(II)). There is a soft repulsion between two particles and between the particles and the boundary, the coefficients of which are presented by \( \kappa \) and \( \kappa_{b} \), respectively \cite{Beppu2021}. Note that we utilize a soft repulsive boundary to prevent confined particles from overlapping at a boundary.

The dynamics of the system is described by a modified version of the Vicsek model \cite{PhysRevLett.75.1226, Beppu2021}. The position of particle \( m \) at time \( t \) is \( \bm{r}_{m}(t) = (x_{m}(t), y_{m}(t)) \), in polar coordinates which becomes \( \bm{r}_{m}(t) = r_m (\cos{\varphi_{m}}, \sin{\varphi_{m}}) \), and the orientation of the particle is \( \bm{d}(\theta_{m}) = (\cos{\theta_{m}}, \sin{\theta_{m}}) \). Each particle has a chirality in motion, which is the inherent tendency to rotate in one direction, denoted by \( \omega \) (Fig.~\ref{fig:fig1}(III)). The equation of motion for the time evolution of the particle position is:
\begin{align}
\Dot{\bm{r}}_m &= \bm{v}(\theta_m(t)) + \frac{2 \kappa}{l^2} \sum_{r_{mn}<\epsilon} (\bm{r}_m - \bm{r}_n) \exp \bigg[ -\bigg( \frac{r_{mn}}{l} \bigg)^2 \bigg] \nonumber\\
&- \kappa_b \hat{\bm{r}}_m \Theta(r_m - R),
\label{eqn_r_eom}
\end{align}
where \( l \) is the length scale of the soft repulsive interaction, \( r_{mn} \) is the distance between particles \( m \) and \( n \), \( \epsilon \) is the radius of polar interaction, and \( \Theta \) is the Heaviside step function, defined as \( \Theta(x) = \{ 1, x>0; 0, x \leq 0 \}\). The equation of motion for the time evolution of the particle orientations is:
\begin{align}
\Dot{\theta}_m &= \omega - \gamma_{p} \sum_{r_{mn}<\epsilon} \sin(\theta_m -\theta_n)  \nonumber\\
&- \gamma_w \sin2 \bigg( \theta_m - \varphi_{m} - \frac{\pi}{2} \bigg) \Theta(r_m - R) + \eta_m,
\label{eqn_theta_eom}
\end{align}
where the random noise \( \eta_{m} \) is related to diffusion coefficient in angle as \( \langle \eta_{m}(t) \eta_{n}(t') \rangle = 2 D \delta_{mn} \delta(t-t') \) (see appendix for implementation details). In dimensionless units, the constant simulation parameters are as follows: \( R=12,\ v_{0}=1,\ \kappa=3,\ \kappa_{b}=20,\ l=0.3,\ \epsilon=1,\ D=0.02 \), which are comparable with data of swimming bacteria in the previous study \cite{Beppu2021}. Simulations were done for a total of at least \( \num{e5} \) time steps, each time step being equal to \( dt = 0.01 \); the differential equations for the particle position and orientation were integrated using the Heun's method.  Longer simulations were performed in conditions where relaxation took longer to confirm whether a steady state had been reached or not. Lastly, the initial positions of the particles are randomly distributed, while the initial orientations are isotropic, with all the particles pointing radially outwards, in order to minimize any initial accidental rotation bias and to investigate the influence of pure chirality on pattern formation.

\section{Results}
\subsection{Achiral active matter \( ( \omega = 0 ) \)}%
\begin{figure*}
\includegraphics[width=0.7\textwidth]{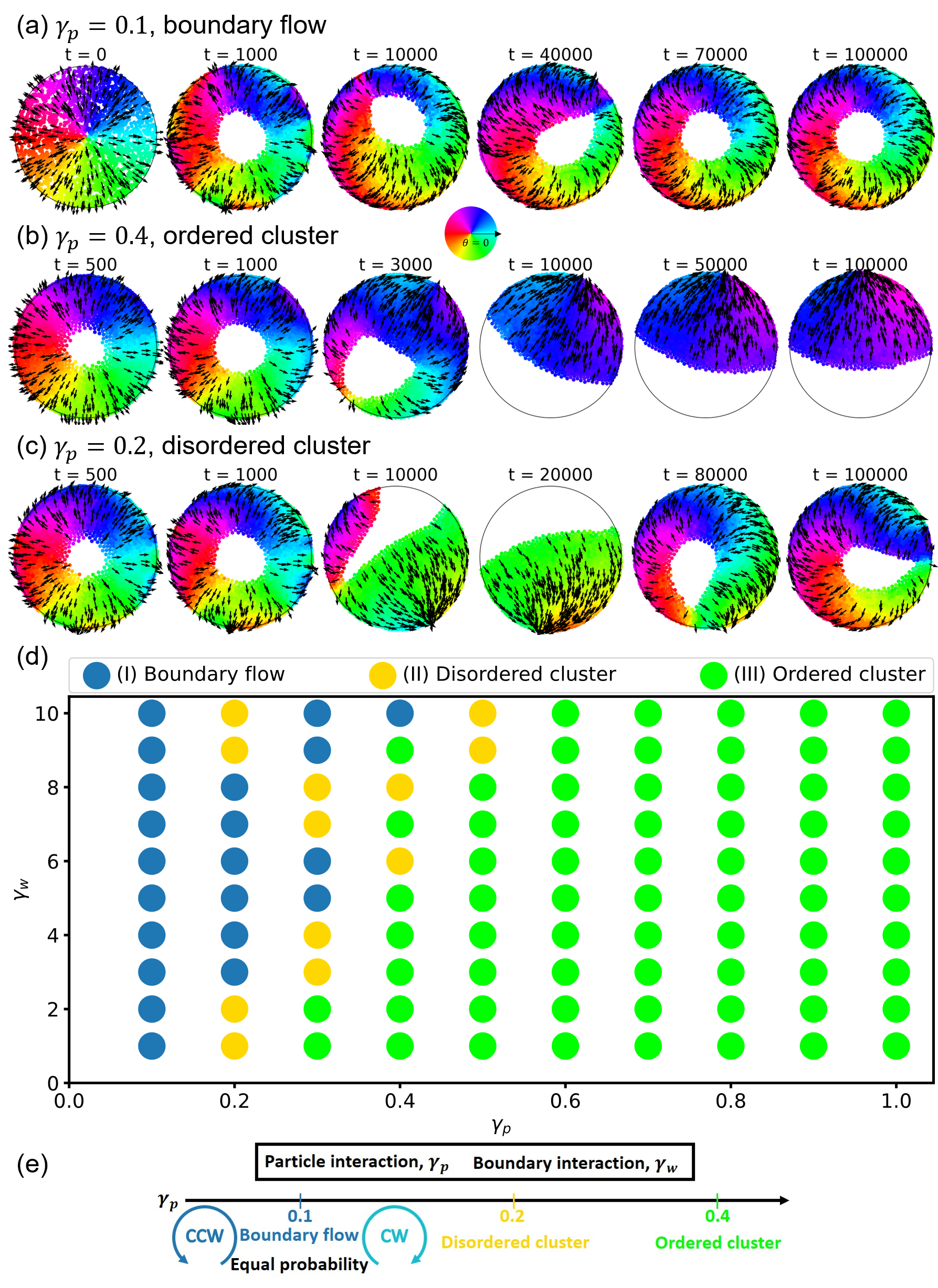}
\caption{
\label{fig:fig2}
\small{Time evolution of high density \((N=3000)\) achiral active matter systems confined to a circular boundary. Depending on the strength of polar (\( \gamma_{p}\)) and nematic (\( \gamma_{w} \)) interactions, achiral systems show one of three possible steady states: (a) boundary flow (\( \gamma_{p} = 0.1, \gamma_{w} = 2\)), (b) ordered cluster (\( \gamma_{p} = 0.4, \gamma_{w} = 2\)) or (c) disordered cluster (\( \gamma_{p} = 0.2, \gamma_{w} = 2\)). Particle color denotes the orientation, arrows denote the velocity vector; for clarity, velocity vectors of 10 percent of total particles shown. (d) Phase diagram for this system; for high density, steady state behavior depends mostly on \( \gamma_{p}\). {\color{black}(e) Summary of phase transitions in achiral systems with \(\gamma_p\). }
}
}
\end{figure*}
This study focuses on the collective motion exhibited by active matter through its interaction with the boundary and its pattern formation. For this aim, we investigate the emergent patterns created by both achiral and chiral confined particles to compare them and clarify the role of chirality and confinement. Firstly, we analyze the collective motion of a group of active matter without chirality and the effect of confinement boundaries. If \( \omega \) is set to \( 0 \) in Eq. \ref{eqn_theta_eom}, the system simplifies to the case of the Vicsek model with excluded volume effects. When such achiral particles are confined within a circular boundary, the system can go to self-organized steady phases depending on the strength of the polar interactions between the particles (\( \gamma_{p} \)) and the nematic interactions with the wall (\( \gamma_{w} \)) (Fig.~\ref{fig:fig2}).

Under conditions of weak polar interactions between particles (\( \gamma_{p} = 0.1\) and \( \gamma_{w} = 2\)), boundary flows are created in which particles move collectively along the boundary (Fig.~\ref{fig:fig2}(a), Video S1). Particles trapped within the circular boundary are oriented tangentially to the boundary and collectively move along the curved wall. Since there is no inherent bias in the motion of the particles (absence of chirality in motion) and no preferred direction after a collision with the boundary, clockwise and counterclockwise boundary flows occur with equal probability. 

By increasing the polar interaction without changing the strength of the interaction with the wall, we examine how the strength of the interaction between the particles changes the nature of the collective motion under a confined space. For \( \gamma_{p} = 0.4\) and \( \gamma_{w} = 2\), particles accumulate in one place, forming a cap-like ordered cluster, which moves very slowly (Fig.~\ref{fig:fig2}(b), Video S2). When the orientation interaction between particles becomes stronger, particles are oriented and move toward each other away from the boundary. Particles accumulate into either a single large, ordered cluster or multiple smaller (typically non-interacting) clusters without being trapped by the boundary because the interaction between particles is stronger than the interaction between particles and the wall. Similar behavior has been observed in previous numerical studies of achiral active matter, in which the particles were found to aggregate at the confining wall \cite{PhysRevE.78.031409, Elgeti_2013}.

Moreover, in the intermediate polar interaction strength (\( \gamma_{p} = 0.2\)  and \( \gamma_{w} = 2\)) between boundary flow (\( \gamma_{p} = 0.1\)) and ordered cluster (\( \gamma_{p} = 0.4\)), a mixed state appears where the cluster structure becomes asymmetric (Fig.~\ref{fig:fig2}(c), Video S3). This fast-moving asymmetric cluster can be thought of as a transition phase between the two other phases, which does not decompose into either of them, even after a long time. Thus, the polar interaction among particles is an essential factor that controls the structure of collective motion in a confined space.

Furthermore, we tested the effect of the nematic interaction with the wall, \( \gamma_{w} \) and drew a phase diagram for those collective motions in \( \gamma_{p} - \gamma_{w} \) (Fig.~\ref{fig:fig2}(d)). The steady state phase is dependent mostly on just \( \gamma_{p} \); below a threshold value of \( \gamma_{p} = 0.3 \), we observe the boundary flow phase (Fig.~\ref{fig:fig2}(d), blue circle) and above it we observe the ordered cluster phase (Fig.~\ref{fig:fig2}(d), lime circle), while the dynamic, mixed state occurs near the threshold \( \gamma_{p} = 0.3 \). The mixed state occurs on the side where \( \gamma_{p} \) is smaller than threshold for small \( \gamma_{w} \) (Fig.~\ref{fig:fig2}(d), yellow circle). This suggests that the interaction \( \gamma_{p} \) between particles must be strong enough to release the interaction with the boundary in order to change the flow along the boundary into a cluster state.

\subsection{Chiral active matter, low \( \gamma_{p}, \gamma_{w} \) regime}
\begin{figure*}
\includegraphics[width=0.7\textwidth]{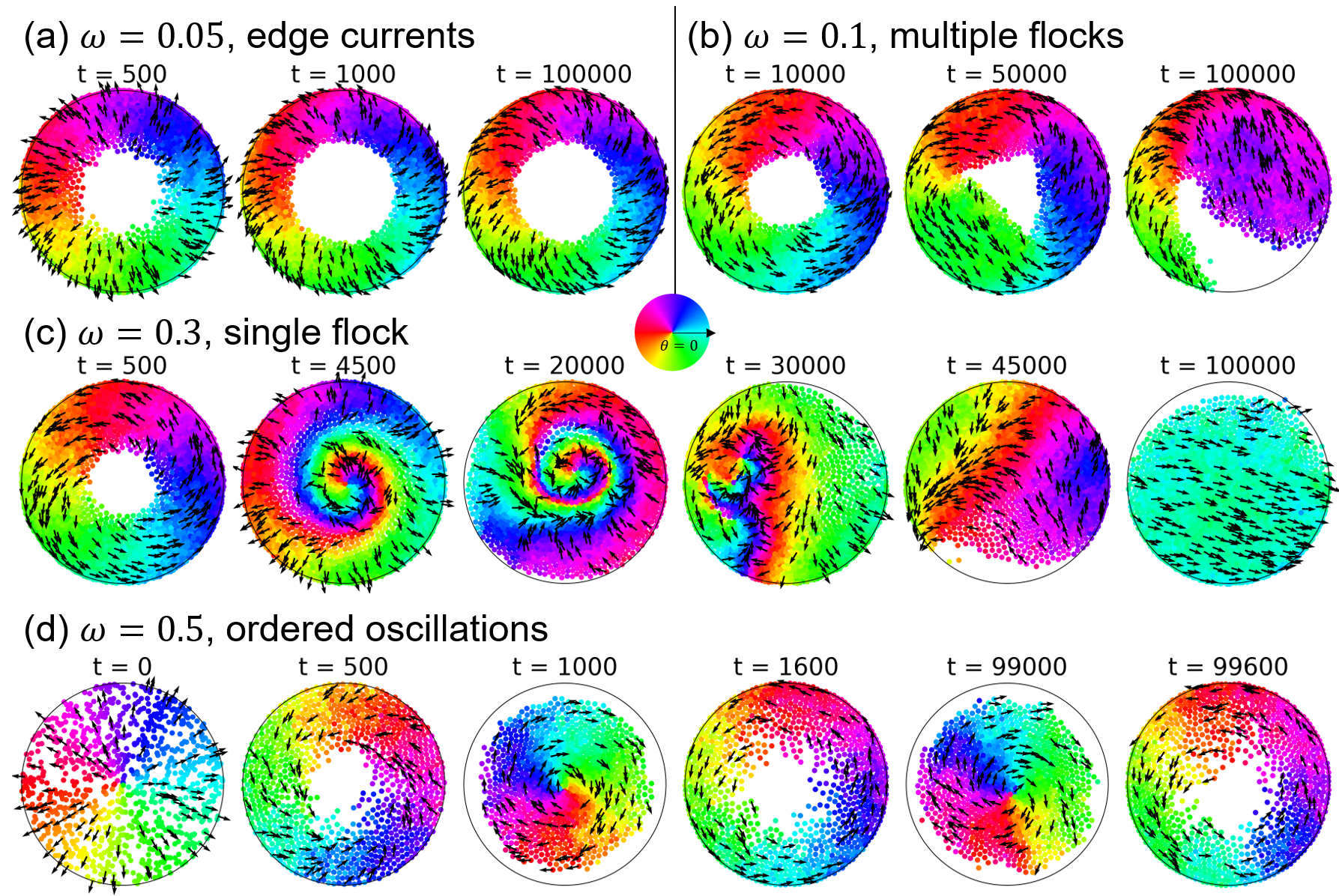}
\caption{
\label{fig:fig3} 
\small{Time evolution of high (a-c, \(N=2000\)) and moderate (d, \(N=1000\)) density chiral active matter systems confined to a circular boundary, with \( \gamma_{p}=0.1,\ \gamma_{w}=1 \). Steady state changes depending on the strength of chirality (\(\omega\)) and particle density (\( N \)): (a) edge currents (\(\omega=0.05\)), (b) multiple flocks (\(\omega=0.1\)), (c) single flock (\(\omega=0.3\)), or (d) ordered oscillations (\(\omega=0.5\)). Particle color denotes the orientation, arrows denote the velocity vector; for clarity, velocity vectors of 10 percent of total particles shown. Representative trajectories of individual particles in each ordered phase shown in Fig.~S2 \cite{supp}.
}
}
\end{figure*}

Next, we focus on the low polar and low nematic interactions regime (\( \gamma_{p}=0.1,\ \gamma_{w}=1 \)) and figure out how the particle density and chirality change the confined collective motion. Since the speed of the particles is constant, the magnitude of \( \omega \) essentially represents the radius of the circular motion (counterclockwise) of an isolated, unconfined particle, and that affects the collective dynamics of chiral particles.

In high-density systems (\( N \geq 1500 \)) at low chirality (\( 0.05 \leq \omega < 0.1\)),  particles start by moving outwards, towards the boundary, and then they are moving along the circular boundary (Fig.~\ref{fig:fig3}(a), Video S4).  The reorientation at the boundary can be in either counterclockwise or clockwise direction, depending on the initial angle of approach of the particles. However, there is an inherent counterclockwise bias to the motion of the particles provided by the chirality, and as the nematic interaction with the wall is weak, reorientation in the clockwise direction is not possible and we only observe these edge currents in the counterclockwise direction.

By raising the magnitude of chirality (\( 0.1 \leq \omega \leq 0.5\)), particles can form either multiple or a single flock, while the effect of chirality can be seen at the onset of each flocking phase. At the moderate chirality, when \( \omega=0.1 \), particles initially move outward towards the boundary and then create an edge current state, but this collective motion later breaks down and forms multiple flocks of particles (Fig.~\ref{fig:fig3}(b), Video S5). By further increasing the chirality bias at (\( 0.1 < \omega \leq 0.5\)), particles initially move outwards towards the boundary and then back towards the center, forming a spiraling structure (Fig.~\ref{fig:fig3}(c), Video S6). However, this spiral flocking state is not stable and only formed transiently, and the group of chiral particles collapses later, leaving the particles to settle down in a flocking steady state. Thus, chirality affects the formation process of ordered phases, and these rotational motions would be important for symmetry breaking.

In addition, because the effect of particle interaction varies with the number density in a confined space, we reduced the number of particles so that the rotational motion of each particle could occur over a wide range and examined how the chiral collective motion changes. Although there is no qualitative difference in the pattern of collective motion that appears when the number of particles is sufficiently large, moderate density systems (\( 200 < N < 1500 \)) show an additional phase. In particular, for the system at the high chirality (\( \omega \geq 0.38 \) for \( N=1000 \)), particles initially move outwards, but quickly turn back, moving towards the center again like an ordered oscillation (Fig.~\ref{fig:fig3}(d), Video S7). The oscillatory state can be found at lower density conditions because the reduced number density allows the particles to be affected by the boundary wall and the polar interaction with neighboring particles. Although the group of particles is trapped in the vicinity of the wall as a boundary flow, the particles can leave the wall due to chiral rotational motion. The particles then gather toward the center, but since the clustered particles at the center rotate according to the chirality, the particles approach the boundary wall again and re-organize into a boundary flow. This oscillatory motion continues for a long time as a periodic change of two states of a flocking and a boundary flow. It is important to note that these ordered oscillations are stabilized by confinement; under periodic boundary conditions without steric constraint, the system initially exhibits oscillatory behavior, but eventually converges to a flocking phase (Fig.~S1, \cite{supp}).

\subsubsection*{Quantitative analysis}
\begin{figure*}
\includegraphics[width=0.9\textwidth]{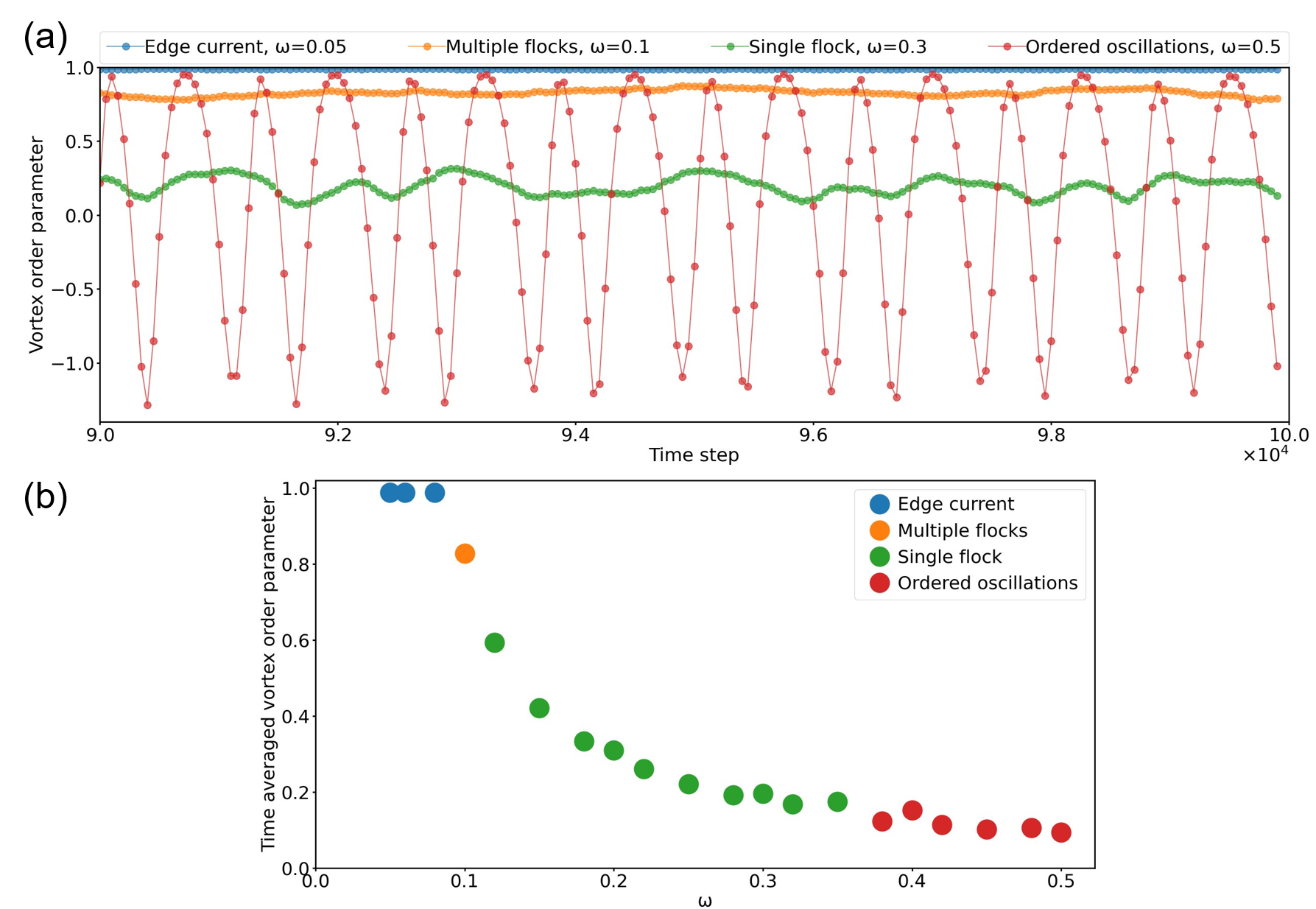}
\caption{
\label{fig:vop_low} 
\small{
For the system with \(N=1000, \gamma_{p}=0.1,\ \gamma_{w}=1\), (a) time evolution of the vortex order parameter (last \(\num{e4}\) time steps) at different steady states: edge currents (blue), multiple flocks (orange), single flock (green), ordered oscillations (red). (b) Time averaged vortex order parameter \(( \langle \Phi_{vop} \rangle )\) versus chirality \(( \omega )\).
}
}
\end{figure*}

\begin{figure*}
\includegraphics[width=0.9\textwidth]{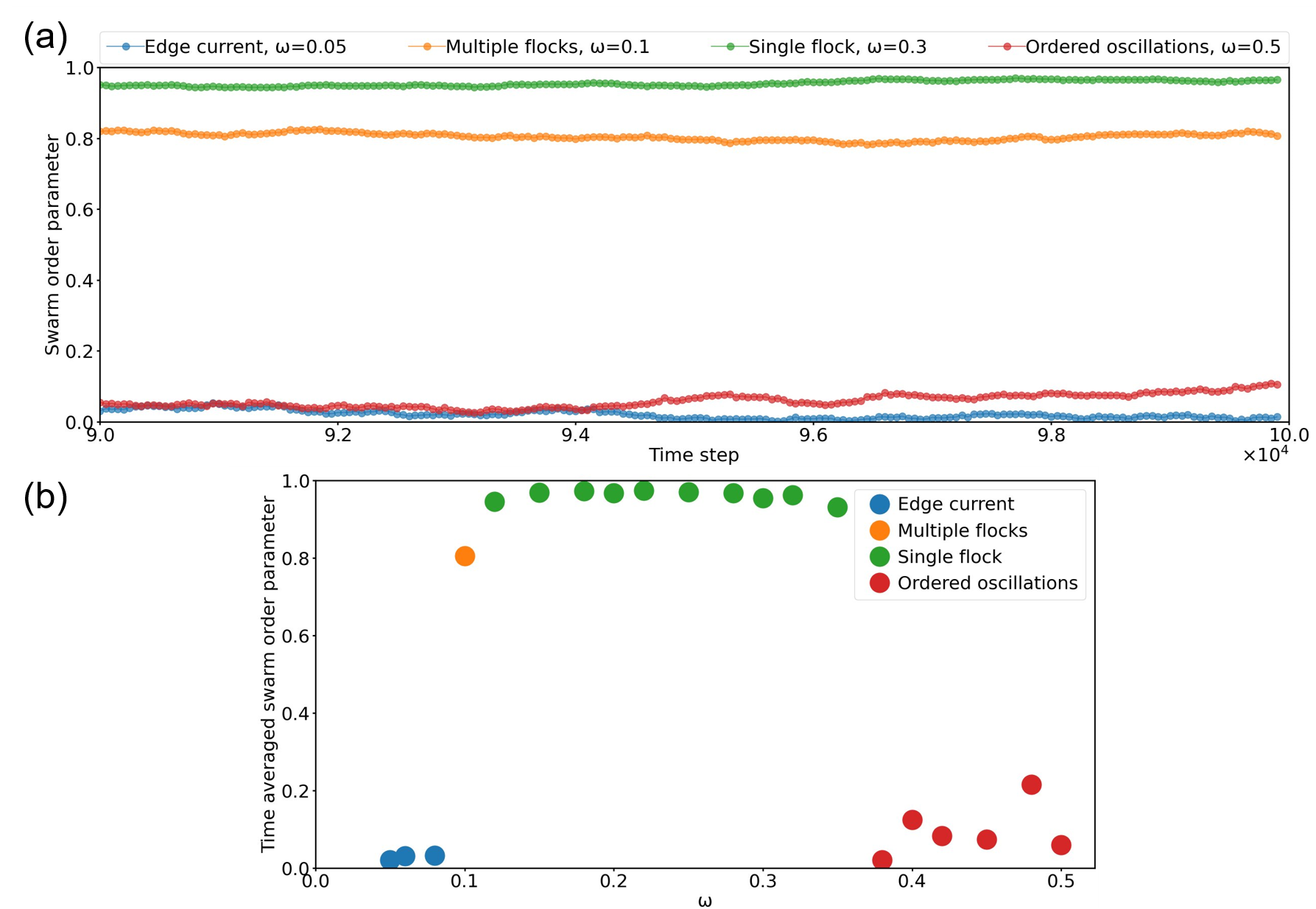}
\caption{
\label{fig:sop_low} 
\small{
For the system with \(N=1000, \gamma_{p}=0.1,\ \gamma_{w}=1\), (a) time evolution of the swarm order parameter (last \(\num{e4}\) time steps) at different steady states: edge currents (blue), multiple flocks (orange), single flock (green), ordered oscillations (red). (b) Time averaged swarm order parameter \(( \langle \psi_{sop} \rangle )\) versus chirality \(( \omega )\).
}
}
\end{figure*}
In a confined region, an ordered pattern, such as a rotating vortex motion along the circular boundary appears in a steady state. On the other hand, a transition from stationary boundary flow to periodic oscillations can be found as the chiral rotation \( \omega \) increases. To demonstrate the difference between the distinct steady states, we performed quantitative analysis extracting characteristics of various ordered phases in this low \( \gamma_{p}, \gamma_{w} \) regime.

The degree of global rotational order can be determined by using the vortex order parameter (VOP, \(\Phi_{vop}\)) \cite{PhysRevLett.110.268102, Beppu2017}. At any time \( t \), VOP is defined as:
\begin{eqnarray}
\Phi_{vop} = \frac{1}{1-2/\pi} \bigg( \frac{\sum_{i} | \bm{v}_{i} \cdot \bm{T}_{i} | }{\sum_{i} || \bm{v}_{i} ||} - \frac{2}{\pi} \bigg),
\label{eqn_vop}
\end{eqnarray}
where \( i \) runs over all the particles, \( \bm{v}_{i} \) is the velocity of particle \( i \), and \( \bm{T}_{i} \) is the unit tangent vector at the position of particle \( i \). \(\Phi_{vop} = 1 \) when the particles move in a perfect vortex, \( \Phi_{vop} = 0 \) when the motion is disordered, and \( \Phi_{vop} < 0 \) when the motion is radial. We calculate the mean VOP, \( \langle \Phi_{vop} \rangle \) by taking the average of the \( \Phi_{vop} \) values for the last \( \num{e4} \) time steps, where the particles have settled into a steady state. Fig.~\ref{fig:vop_low}(a) shows the time evolution of \( \Phi_{vop} \) for moderate (\( N=1000 \)) density systems, and Fig.~\ref{fig:vop_low}(b) shows how \( \langle \Phi_{vop} \rangle \) varies with \(\omega\). Higher density systems show the same trends, except that the ordered oscillation phase is absent {\color{black}(Fig.~S21)}.

We found that the edge current phase has a practically constant \( \Phi_{vop} \); for \( \omega=0.05 \), \( \langle \Phi_{vop} \rangle = 0.99 \) (Fig.~\ref{fig:vop_low}(b), blue), indicating a highly ordered vortex phase. In addition, the multiple flocks phase also has an almost constant \( \Phi_{vop} \), its magnitude being slightly less than \( 1 \); for \( \omega=0.1 \), \( \langle \Phi_{vop} \rangle = 0.83 \) (Fig.~\ref{fig:vop_low}(b), orange). The slightly lower value in this phase compared to the edge current phase reflects that the entire system of particles does not move along the boundary but is oriented radially due to the distorted shape of the multiple flocks. A similar pattern occurs in the single flock phase, where \( 0 < \Phi_{vop} < 0.6\); for \( \omega=0.3 \), \( \langle \Phi_{vop} \rangle = 0.20 \) (Fig.~\ref{fig:vop_low}(b), green), and \( \langle \Phi_{vop} \rangle \) decreases as \(\omega\) increases. A transition from stationary edge current to periodic oscillations can be found as the chiral rotation \( \omega \) increases for moderate density systems. For the ordered oscillation phase, \( \Phi_{vop} \) shows periodic oscillation between 1 and -1.3; for \( \omega=0.5 \), \( \langle \Phi_{vop} \rangle = 0.09 \) (Fig.~\ref{fig:vop_low}(b), red). In this phase, particles obtain a radial velocity  between the wall side and the center of confined space, and the periodic repetition of this radial motion and the motion along the wall gives oscillatory change of \( \Phi_{vop} \) over time. Furthermore, since \(\Phi_{vop}\) goes from its maxima to its minima twice for each oscillation of the system, the angular frequency \((\omega_{\Phi_{vop}})\) of the oscillating \( \Phi_{vop} \) is approximately twice the chirality \((\omega)\), indicating that the global motion of the system corresponds to the individual motion of the particles; for \( \omega=0.5 \), \( \omega_{\Phi_{vop}} = 1.01 \). The angular frequency obtained from considering the periodic motion of the particles however, is roughly equal to the chirality (Fig.~S3). Moreover, the ordered oscillations state can occur for other initial orientations, as long as the particles do not get aligned into a flock (Fig.~S4).

On the other hand, the highly ordered motion of the flocking phase can be quantified through the swarm order parameter (SOP, \(\psi_{sop}\)) \cite{D2SM01402E}, which is defined as:
\begin{eqnarray}
{
\color{black}
\psi_{sop} = \frac{1}{N} \sum_{i} \cos(\theta_i - \bar{\theta}),
}
\label{eqn_sop}
\end{eqnarray}
where \( i \) runs over all the particles, \(N\) is the total number of particles, {\color{black} \(\theta_i\) is the orientation of particle \(i\) and \( \bar{\theta} \) is the mean orientation angle of all the particles obtained by calculating $\bar{\theta} = \tan^{-1}\bigl(\frac{\sum_{i} \sin\theta_i}{\sum_{i}\cos\theta_i}\bigr)$ \cite{D2SM01402E}}. When all the particles are aligned in the same direction, \( \psi_{sop} = 1 \), whereas if their orientations are in different directions, \( \psi_{sop} = 0 \). The mean SOP, \( \langle \psi_{sop} \rangle \) is also calculated in the same manner as \( \langle \Phi_{vop} \rangle \). For moderate (\( N=1000 \)) density systems, the time evolution of \( \psi_{sop} \) is shown in Fig.~\ref{fig:sop_low}(a), and the dependence of \( \langle \psi_{sop} \rangle \) on \(\omega\) in Fig.~\ref{fig:sop_low}(b).

As expected, \( \psi_{sop} \) remains close to 0 for the edge current and the ordered oscillation phases, in which the particles are oriented in all directions; \( \langle \psi_{sop} \rangle = 0.02\) for \(\omega=0.05\) (Fig.~\ref{fig:sop_low}(b), blue), and \( \langle \psi_{sop} \rangle = 0.06\) for \(\omega=0.5\) (Fig.~\ref{fig:sop_low}(b), red). For the multiple flocks phase, \( \psi_{sop} \) stays slightly lower than 1 whereas for the single flock phase, it becomes very close to 1; \( \langle \psi_{sop} \rangle = 0.80\) for \(\omega=0.1\) (Fig.~\ref{fig:sop_low}(b), orange), and \( \langle \psi_{sop} \rangle = 0.96\) for \(\omega=0.3\) (Fig.~\ref{fig:sop_low}(b), green). Note that, for the single flock phase, \( \langle \psi_{sop} \rangle \) increases as \(\gamma_p\) increases, but \( \langle \psi_{sop} \rangle \) stays almost constant with \(\gamma_w\) (Fig.~S5).

By using \( \Phi_{vop} \) and \( \psi_{sop} \), we can quantitatively define all the dynamic steady states, as described in the appendix. Furthermore, we can use other order parameters such as the dynamics of the center of mass (Fig.~S6), mean radius (Fig.~S7), variance of radius (Fig.~S8), and mean tangential velocity (Fig.~S9) to obtain more information about the spatial distribution and dynamics of the different phases.

\subsubsection*{Phase diagram}
\begin{figure*}
\includegraphics[width=0.95\textwidth]{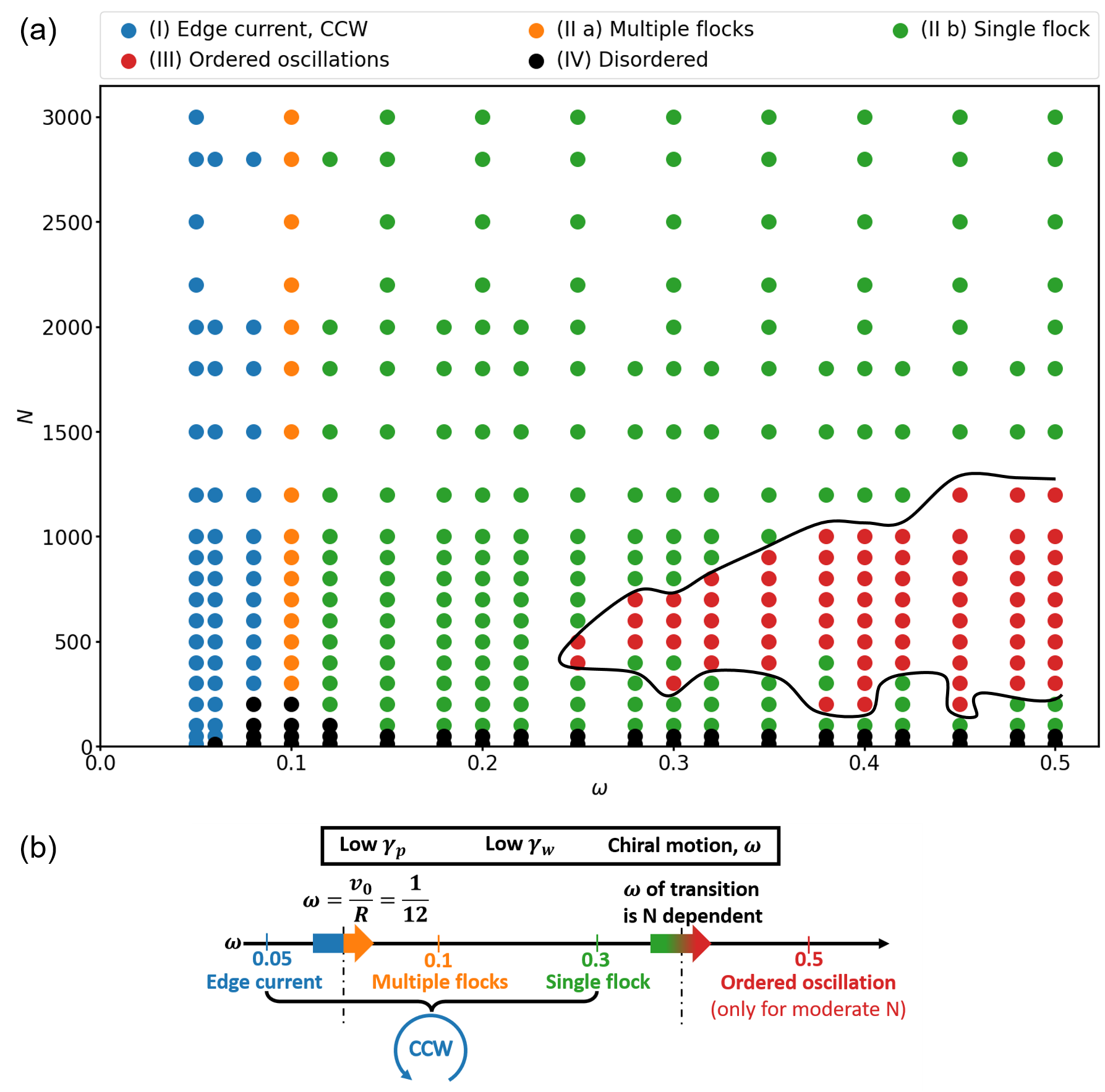}
\caption{
\label{fig:fig5} 
\small{(a) Phase diagram in the low polar and nematic interactions regime (\( \gamma_{p}=0.1, \gamma_{w}=1 \)). Above a minimum threshold particle density, for low \( \omega \), we observe counterclockwise edge currents; by increasing \( \omega \), the systems phase transitions to flocking, and for moderate density systems, further increasing \( \omega \) leads to the ordered oscillation phase. Phase boundaries are manually drawn and are just a guide for the eye. (b) Summary of how the phase changes with chirality \(( \omega )\).
}
}
\end{figure*}
To summarize the transitions with chirality and number density, we plot a phase diagram for this low \( \gamma_{p}, \gamma_{w} \) regime (Fig.~\ref{fig:fig5}(a)), with the chirality (\( \omega \)) on the \( x \) axis and the number of particles (\( N \)) on the \( y \) axis.

For low \( \omega \), irrespective of the particle density, we observe an edge current phase, where the particles move along the circular boundary. Due to the inherent bias provided by the chirality, only counterclockwise edge currents are observed in this regime. The systems start transitioning from the edge current phase at \( \omega = v_{0}/R = 1/12 \simeq 0.08 \) irrespective of the density. As  chirality is increased to moderate \( \omega \) values, all the systems are in a single flock phase. The transition from the edge current phase to the single flock phase always happens through an intermediate multiple flocks phase.

On the other hand, for high \( \omega \), systems with moderate density (\( 200 < N < 1500 \)) show an ordered oscillation phase. However, the minimum \( \omega \) required to induce this phase depends on the particle density. For \( N=500 \), the minimum \( \omega \) is \(0.25\), and this minimum required \( \omega \) increases for systems with \( N < 500 \) as well as \( N > 500 \), and hence this transition shows a nonlinear dependence on \( N \) and \( \omega \). The phase dependence on \(\omega\) and \(N\) is summarized in Fig.~\ref{fig:fig5}(b).

We note that below a minimum threshold particle density (\( N \leq 200 \)), the particle distribution can be too sparse to show any sort of collective motion, and their dynamics does not change from a disordered state. In contrast, high density (\( N \geq 1500 \)) systems do not show the ordered oscillation phase at all and show flocking even for high \( \omega \). This is because if the density is too large, the inter-particle interaction becomes dominant and the effective interaction with the wall is relatively weakened.

\subsection{Chiral active matter, high \( \gamma_{p}, \gamma_{w} \) regime}
\begin{figure*}
\includegraphics[width=0.7\textwidth]{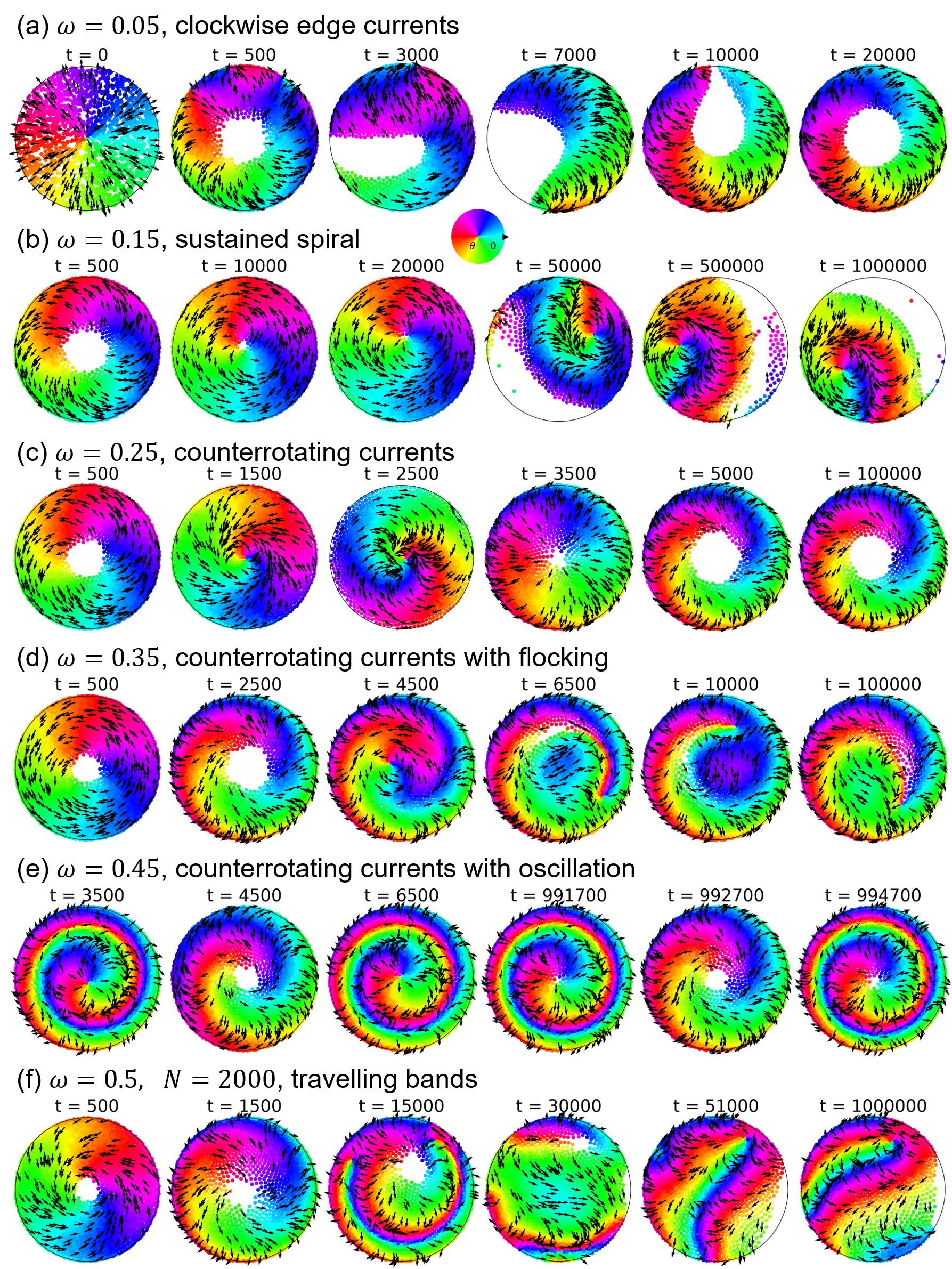}
\caption{
\label{fig:fig6} 
\small{Time evolution of chiral active matter systems confined to a circular boundary at high density ((a-e) \(N=2500\), (f) \(N=2000\)), with \( \gamma_{p}=0.5,\ \gamma_{w}=10 \). In this regime, different steady states from before are observed: (a) clockwise edge currents (\(\omega=0.05\)), (b) sustained spiral (\(\omega=0.15\)), (c) counterrotating currents (\(\omega=0.25\)), (d) counterrotating currents with flocking (\(\omega=0.35\)), (e) counterrotating currents with oscillation and (f) travelling bands (\(\omega=0.5\)). Particle color denotes the orientation, arrows denote the velocity vector; for clarity, velocity vectors of 10 percent of total particles shown. Representative trajectories of individual particles in each ordered phase shown in Fig.~S10 \cite{supp}.
}
}
\end{figure*}
Till now, we have been focusing on the low \( \gamma_{p}, \gamma_{w} \) regime. In the absence of chirality, this regime favours the boundary flow phase (Fig.~\ref{fig:fig2}(a)), while the high \( \gamma_{p}, \gamma_{w} \) regime favours clustering (Fig.~\ref{fig:fig2}(b)). In following sections, we examine how chirality affects this clustering regime.

We perform the simulations with the same parameters as given in earlier sections, except that the polar interparticle interactions and nematic interactions at the boundary are stronger, \( \gamma_{p}=0.5,\ \gamma_{w}=10 \) respectively. In this regime, collective motion previously unseen are observed at high particle density (Fig.~\ref{fig:fig6}, \(N=2500\)(a-e), 2000(f)).

In this regime, clockwise edge currents are possible at low chiralities (Fig.~\ref{fig:fig6}(a) and  Video S8, \( \omega = 0.05 \)). Particles start by moving towards the boundary, and as the nematic interaction with the boundary is strong, if the reorientation is in the clockwise direction, the strong alignment along the boundary can dominate over the polar particle interaction (\( t=0-500 \)). We observed a group of particles with clockwise alignment at the boundary along with the counterclockwise aligned particles, and thus a disordered cluster formed (\( t=3000 \)). However, this transient state immediately decomposes into a clockwise edge current phase (\( t=7000-10000 \)). The particles not directly in contact with the boundary can also move in the clockwise direction (\( t=20000 \)) because the relatively higher polar interaction between the particles is able to propagate the clockwise alignment at the boundary inwards, which is able to dominate over the comparatively small chirality. Such clockwise edge currents were absent in the low \( \gamma_{p}, \gamma_{w} \) regime, and demonstrate that nematic interactions with the boundary are capable of overpowering the inherent chirality of the particles. For low chirality, the system can go to the clockwise edge current steady state, irrespective of confinement size (Fig.~S11) and initial orientations (Fig.~S12). It may also go to a counterclockwise edge current or a disordered cluster, however, the probability of the states does depend on the initial conditions (Figs.~S13 and S14). For the system with radially outward initial orientations shown in Fig.~\ref{fig:fig6}(a), at very low chiralities the disordered cluster is slightly more probable, but as \(\omega\) increases, we primarily observe counterclockwise edge currents. Interestingly, the speed of the clockwise edge current phase is significantly less than that of the counterclockwise one (Fig.~S15). This happens because the counterclockwise chirality opposes clockwise motion due to the strong nematic interaction.

By increasing chirality, the sustained spiral pattern was observed (Fig.~\ref{fig:fig6}(b) and Video S9, \( \omega = 0.15 \)). Particles first move towards the boundary (\( t=500 \)), then back towards the center, forming a symmetric spiral structure covering the whole confinement area (\( t=10000-20000 \)). The spiral continues for a while but no longer stays symmetric and forms a spiraling droplet that itself moves along the circular boundary in a steady state (\( t=50000 \)). At one of the points of intersection of the spiral with the boundary, we can find that the particles on one side are aligned in the clockwise direction, and on the other side, they are aligned in the counterclockwise direction, and the continuous interaction of these particles acts as a feedback loop and is the reason we observe this sustained spiral.

For moderate chirality, collective motion of particles form counterrotating currents (Fig.~\ref{fig:fig6}(c) and Video S10, \(\omega = 0.25\)). The particles initially form a spiralling droplet (\( t=500-1500 \)), and when the particles move outwards to the boundary again, the outermost particles get aligned tangent to the boundary in the clockwise direction, and due to the strong nematic interaction with the wall (\( t=2500-3500 \)), they are able to keep that alignment and hence clockwise currents appear near the boundary. However, the particles far from the boundary are mostly dependent on the polar interaction between the particles to propagate the clockwise alignment inwards from the boundary (\( t=5000 \)). In this moderate chirality regime, the polar interaction between the particles cannot dominate over the chirality. The particles away from the boundary move in the counterclockwise direction, thus forming counterrotating currents.

By raising the chirality a little higher, the counterrotating current with flocking phase was observed (Fig.~\ref{fig:fig6}(d) and Video S11, \(\omega = 0.35\)). The system initially has the same behavior as the counterrotating currents (\( t=0-2500 \)), but since \( \omega \) is higher, the radius of the particles' circular motion is smaller, and thus the particles further away from the boundary break away from the counterrotating currents (\( t=6500 \)) and perform flocking around the center of the boundary (\( t=10000 \)).

In contrast, for high chirality, counterrotating currents started to show oscillation (Fig.~\ref{fig:fig6}(e) and Video S12, \(\omega = 0.45\)). The particles initially have a similar behavior as the counterrotating currents, but as the chirality is very high, a spiral forms near the center of the circle (\( t=3500 \)). Similar to the ordered oscillation phase in the low \( \gamma_{p}, \gamma_{w} \) regime, this spiral also shows oscillatory formation and deformation (\( t=4500-6500 \)).

We note that a travelling band pattern was observed for high chirality but at slightly lower densities (Fig.~\ref{fig:fig6}(f) and Video S13, \( N = 2000 \), \(\omega = 0.5\)). Since the chirality is high, the system initially seems to be in the counterrotating currents with oscillations (\( t=1500 \)), but the oscillations in this case soon become unstable and the phase collapses (\( t=15000 \)) in such a way that some of the particles form a flock that circles around in the inside of the confinement and some of the particles form a disordered cluster at the boundary (\( t=30000 \)). When the flock encounters the cluster, chiral particles form travelling bands (\( t=51000 \)).

We also note that, similar to ordered oscillations in the low \( \gamma_{p}, \gamma_{w} \) regime, the sustained spiral, the travelling waves and all of the counterrotating currents are stabilized due to the confinement, without which these systems form a flocking steady state instead (see Supplemental Material \cite{supp}, Fig.~S1).

\subsubsection*{Quantitative analysis}
\begin{figure*}
\includegraphics[width=0.9\textwidth]{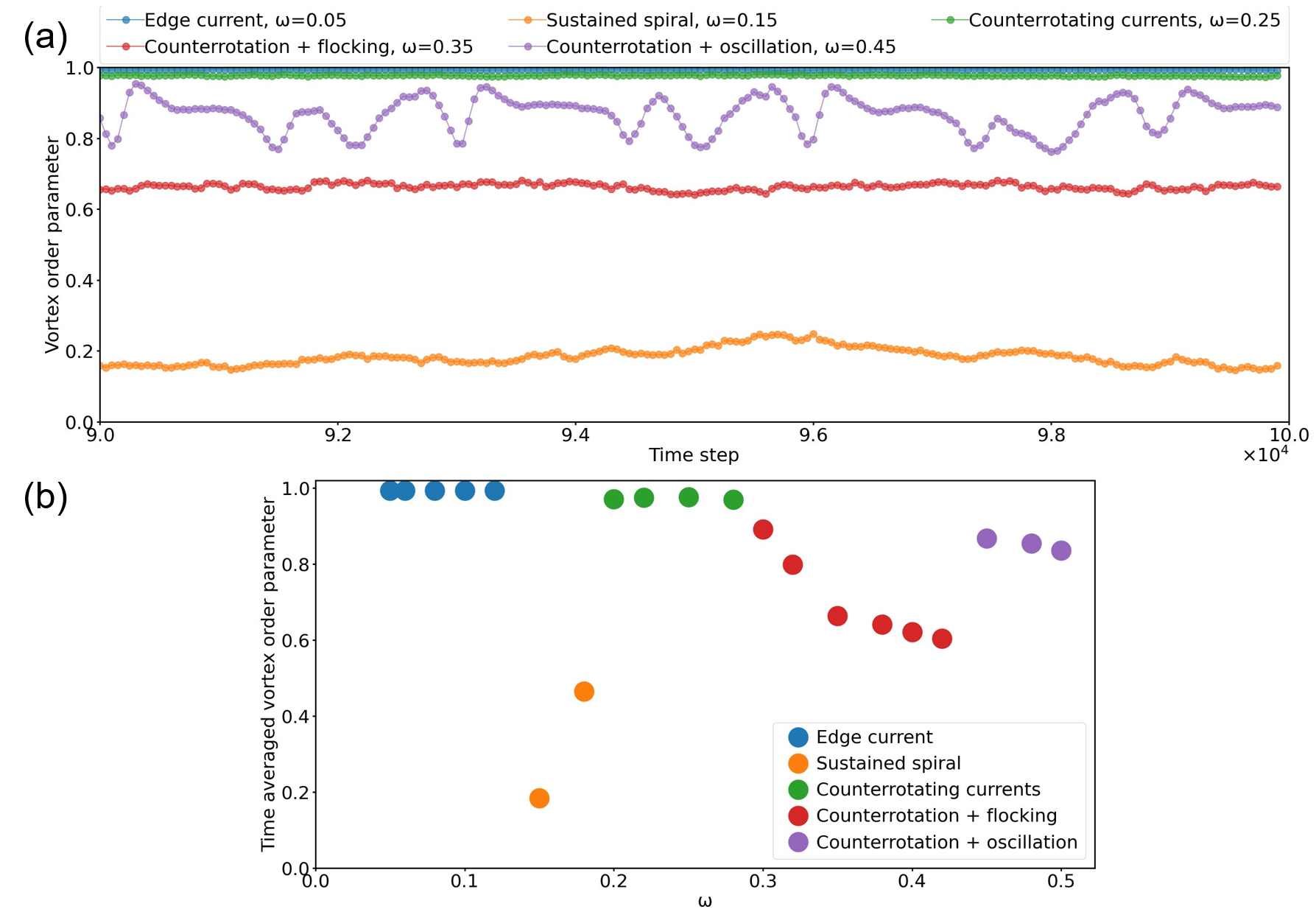}
\caption{
\label{fig:vop_high} 
\small{
For the \(N=2500, \gamma_{p}=0.5,\ \gamma_{w}=10\) system, (a) time evolution of the vortex order parameter (last \(\num{e4}\) time steps) for edge currents (blue), sustained spiral (orange), counterrotating currents (green), counterrotations with flocking (red), counterrotations with oscillation (purple). (b) Time averaged vortex order parameter \(( \langle \Phi_{vop} \rangle )\) versus chirality \(( \omega )\).
}
}
\end{figure*}
\begin{figure*}
\includegraphics[width=0.9\textwidth]{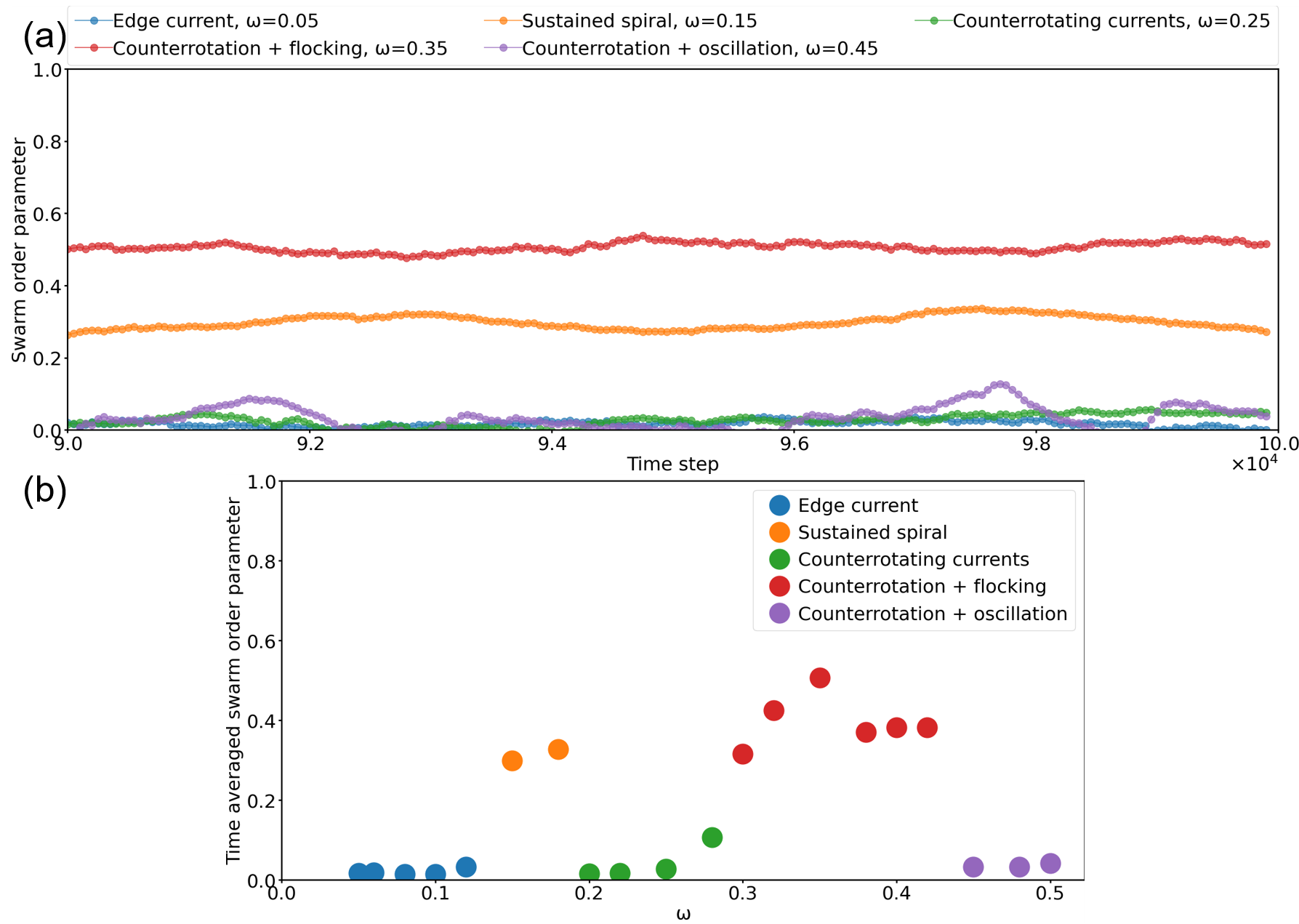}
\caption{
\label{fig:sop_high}
\small{
For the  \(N=2500, \gamma_{p}=0.5,\ \gamma_{w}=10\) system, (a) time evolution of the swarm order parameter (last \(\num{e4}\) time steps) for edge currents (blue), sustained spiral (orange), counterrotating currents (green), counterrotations with flocking (red), counterrotations with oscillation (purple). (b) Time averaged swarm order parameter \(( \langle \psi_{sop} \rangle )\) versus chirality \(( \omega )\). {\color{black}We note that the swarm order parameter is evaluated by only considering particles that are within a distance of 9 from the center of mass (\(\mathbf{r}_m\)) of the particles.}
}
}
\end{figure*}
We do quantitative analysis for the high density systems in this high \( \gamma_{p}, \gamma_{w} \) regime. We consider again the VOP and SOP, as defined earlier (Eq. \ref{eqn_vop} and Eq. \ref{eqn_sop}). Fig.~\ref{fig:vop_high}(a) shows the time evolution of \(\Phi_{vop}\) different chiralities and Fig.~\ref{fig:vop_high}(b) shows \(\langle \Phi_{vop} \rangle\) versus \( \omega \) for \(N=2500\).

Same as before, \( \Phi_{vop} \) in the edge current phase is nearly constant and almost equal to 1; for \( \omega=0.05 \), \( \langle \Phi_{vop} \rangle = 0.99 \) (Fig.~\ref{fig:vop_high}(b), blue). Comparatively, \( \Phi_{vop} \) becomes small for the sustained spiral phase; for \( \omega=0.15 \), \( \langle \Phi_{vop} \rangle = 0.18 \) (Fig.~\ref{fig:vop_high}(b), orange). For the counterroating currents phase, \( \Phi_{vop} \) is just marginally smaller than the edge currents phase; for \( \omega=0.25 \), \( \langle \Phi_{vop} \rangle = 0.98 \) (Fig.~\ref{fig:vop_high}(b), green). This is because some particles between the clockwise rotating boundary layer and the counterclockwise rotating innermost layer are aligned in the radially outward direction, and hence do not contribute to the average. \( \Phi_{vop} \) decreases further in the counterrotation with flocking phase, it being  \( 0.6 < \Phi_{vop} < 0.9\); for \( \omega=0.35 \), \( \langle \Phi_{vop} \rangle = 0.66 \) (Fig.~\ref{fig:vop_high}(b), red). The flocking reduces the rotational order, and similar to the low \(\gamma_p, \gamma_w\) regime, as \(\omega\) increases, \( \langle \Phi_{vop} \rangle \) decreases. For the counterrotation with oscillation phase, similar to the ordered oscillation phase in the low \(\gamma_p, \gamma_w\) regime, \( \Phi_{vop} \) shows periodic oscillation between 0.7 and 1; for \( \omega=0.45 \); \( \langle \Phi_{vop} \rangle = 0.84 \) (Fig.~\ref{fig:vop_high}(b), purple). The angular frequency \(( \omega_{\Phi_{vop}} )\) of the oscillating \(\Phi_{vop}\) in this phase is also approximately two times the chirality \((\omega)\); for \( \omega=0.45 \), \( \omega_{\Phi_{vop}} = 0.88 \).

To get a better picture of the flocking behavior in this regime, \( \psi_{sop} \) is calculated in a limited area around the center of mass \((\mathbf{r}_m)\) of the particles (within a distance of 9 from \(\mathbf{r}_m\)). For the \( N=2500 \) system, the time evolution of \( \psi_{sop} \) is shown in Fig.~\ref{fig:sop_high}(a), and the dependence of \( \langle \psi_{sop} \rangle \) on \(\omega\) in Fig.~\ref{fig:sop_high}(b). Just like the low \( \gamma_p, \gamma_w \) regime, \( \psi_{sop} \) remains close to 0 for the edge current, the counterrotating currents and the counterrotation with oscillation phases; \( \langle \psi_{sop} \rangle = 0.02\) for \(\omega=0.05\) (Fig.~\ref{fig:sop_high}(b), blue), \( \langle \psi_{sop} \rangle = 0.03\) for \(\omega=0.25\) (Fig.~\ref{fig:sop_high}(b), green) and \( \langle \psi_{sop} \rangle = 0.04\) for \(\omega=0.45\) (Fig.~\ref{fig:sop_high}(b), purple). The sustained spiral and the counterrotation with flocking phases show intermediate values of \( \psi_{sop} \), between 0.2 and 0.6; \( \langle \psi_{sop} \rangle = 0.3\) for \(\omega=0.15\) (Fig.~\ref{fig:sop_high}(b), orange), and \( \langle \psi_{sop} \rangle = 0.5\) for \(\omega=0.35\) (Fig.~\ref{fig:sop_high}(b), red). The quantitative definition of every state using the order parameters is given in the appendix. Also, same as the low \(\gamma_p, \gamma_w\) regime, the center of mass dynamics (Fig.~S16), mean radius (Fig.~S17), variance of radius (Fig.~S18), and mean tangential velocity (Fig.~S19) provide us with more information about the spatial distribution and dynamics of the different phases in this regime as well. Quantitative analysis for the travelling bands and the disordered cluster phases are also given in Fig.~S20.

\subsubsection*{Phase diagram}
\begin{figure*}
\includegraphics[width=0.95\textwidth]{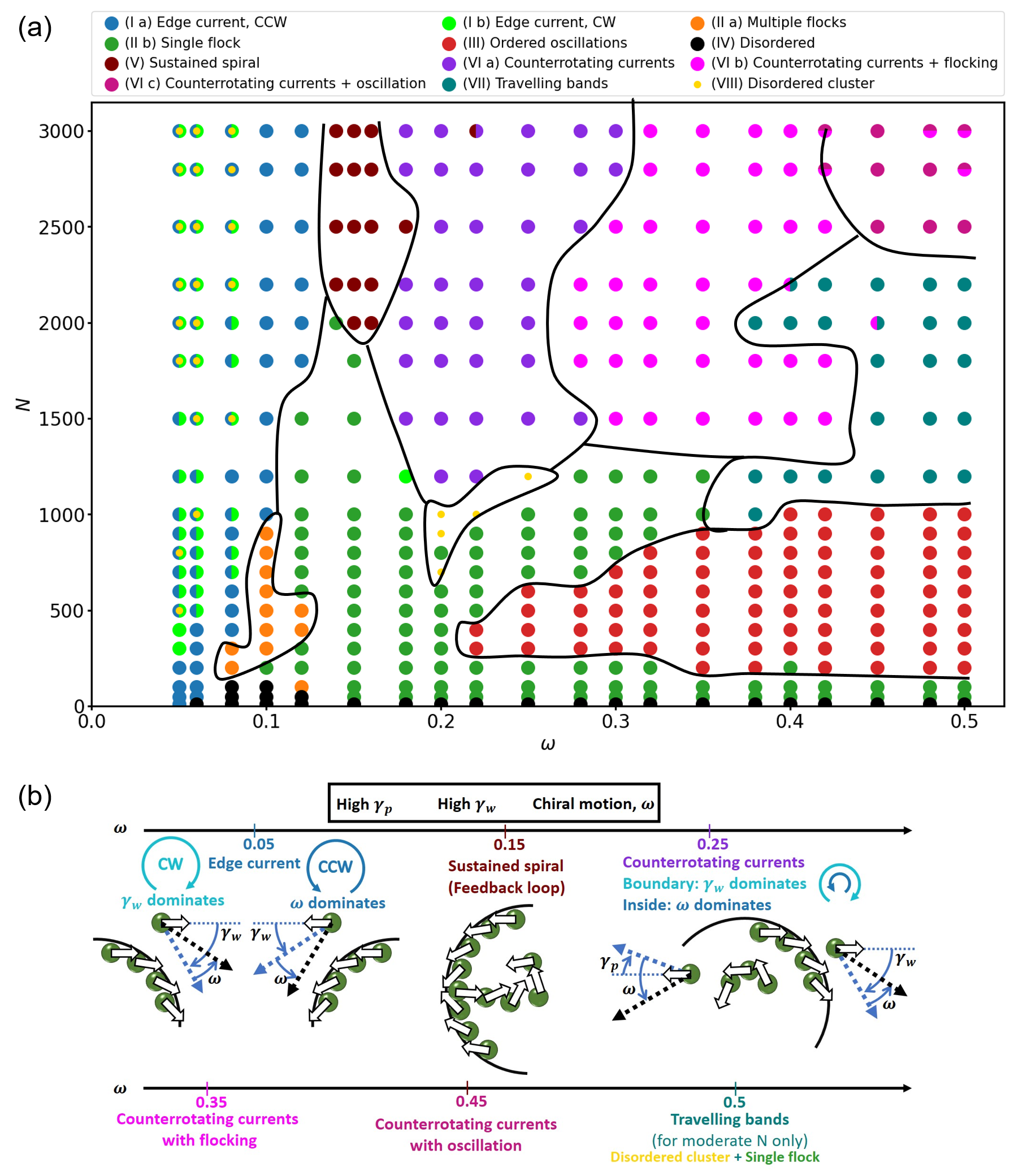}
\caption{
\label{fig:fig8}
\small{(a) Phase diagram in the high polar and nematic interactions regime (\( \gamma_{p}=0.5, \gamma_{w}=10 \)). Compared to the low \( \gamma_{p}, \gamma_{w} \) regime, the high density systems show much more complex steady states. Points with different colors on left and right halves of the circle represent different simulation runs with the same initial conditions can give either one of the states. Some of the counterrotating currents with oscillation systems show the oscillations for some time, but then oscillations collapse to flocking; these are represented by points with different colors on top and bottom halves of the circle. Phase boundaries are manually drawn and are just a guide for the eye. (b) Summary of the phase changes with chirality \(( \omega )\) and schematic diagrams for the steady states. 
}
}
\end{figure*}
Finally, we summarize all the simulation data in the high \( \gamma_{p}, \gamma_{w} \) regime in a phase diagram (Fig.~\ref{fig:fig8}(a)), with the chirality (\( \omega \)) and number of particles (\( N \)) on the \(x\) and \(y\) axes respectively. Same as the low \( \gamma_{p}, \gamma_{w} \) regime, below a minimum threshold particle density (\( N \leq 200 \)), we can observe a disordered state without any collective motion. Moreover, the edge current phase exists for low chirality across all particle densities. However, owing to the strong nematic interaction with the boundary, if a significant number of particles get aligned in the clockwise direction, it is possible for clockwise edge currents to appear as well.  Furthermore, likely due to the higher polar interactions between the particles, in systems with moderate and high densities (\( N \geq 1000 \)), the edge current phase is also observed for slightly higher chiralities (\( \omega =0.1, 0.12 \)) compared to the low \( \gamma_{p}, \gamma_{w} \) systems, replacing the multiple flocks phase.

A few systems with high density/low chirality and moderate density/moderate chirality settle down in to the disordered cluster phase. The achiral (\(\omega=0\)) systems with \( \gamma_{p}=0.5, \gamma_{w}=10 \), are either in the ordered or disordered cluster phase. When a few particles are aligned in the clockwise direction, they compete with the clockwise motion due to chirality to determine the global motion of the system, and if neither is able to completely overcome the other, we observe the disordered cluster phase.

In contrast, for very low chirality, the system may either have a counterclockwise edge current, clockwise edge current or a disordered cluster steady state, depending on whether the chirality, the nematic interaction or neither dominates, and that might be different from one simulation run to another, even if none of the parameters are changed. Thus, we observe multiple steady states for the low chirality systems.

It is in the high density and moderate to high chirality region of the phase diagram that we observe novel and interesting phases, in stark contrast to the low \( \gamma_{p}, \gamma_{w} \) regime, which had just flocking for that region. The sustained spiral phase is a peculiar state that occurs for a small and specific range of parameters: \( N \geq 2000 \) and around \( \omega = 0.15 \); transient spiralling droplets are observed for a wide range of parameters (the single flock phase has an intermediate spiraling droplet state, in both low and high \( \gamma_{p}, \gamma_{w} \) regimes), but within this small subset of parameters, the conditions are just right to observe long-lasting spiralling droplets.

On the other hand, increasing the chirality beyond the sustained spiral phase leads us to the counterrotation phases, in which the particles close to the boundary move in the clockwise direction (due to strong nematic interaction with the wall), while the particles further away from the boundary move in the counterclockwise direction (due to the higher chirality). On the higher end of \( \omega \), the counterrotations are also accompanied with either flocking or an oscillatory spiralling motion of particles at the center of the circular boundary. The phase dependence on \(\omega\) and \(N\) in this regime, is summarized in Fig.~\ref{fig:fig8}(b), along with the schematics for the steady states.

\section{Discussion}
In this study, we numerically studied the pattern formation due to the collective motion of chiral active matter confined in a circular boundary space. Our simulations show that introducing chirality into a confined achiral active matter system can drastically change its dynamics; while the achiral systems show one of the three simple states of boundary flow, ordered cluster or a mixed state, the chiral systems, depending on the particle number \((N)\), chirality \( ( \omega ) \), polar \( ( \gamma_{p} ) \)  and nematic \( ( \gamma_{w} ) \) interactions show a wide variety of steady states. For chiral systems in the low \( \gamma_{p}, \gamma_{w} \) regime, we observe chiral edge currents, flocking, and ordered oscillations, whereas in the high \( \gamma_{p}, \gamma_{w} \) regime we observe additional phases that are much more dynamic: counterrotations (with its variants), sustained spirals, and traveling bands.

The particles interact with a boundary wall as they move in a curved trajectory with a preferential direction. Under conditions where the interaction with the boundary wall and the polar orientation interaction between particles are weak, a boundary flow along the wall emerges at lower chirality, and as the chirality is increased, the chiral collective motion transforms into a global oscillation that reverses the direction in the radial direction within the confined space. Such ordered collective motion suggests the presence of an effective attractive interaction through the chiral motion of particles near the boundary. Furthermore, as the strength of the interaction between the particles and the wall becomes stronger, various order formations, such as counterrotations, are stabilized according to the balance of interparticle alignment and steric effect at the boundary. Thus, symmetry breaking of the self-propelled particles, which could be coupled with the steric boundary condition, is a critical parameter controlling the macroscopic collective dynamics.

The emergence of different states when achiral systems are transformed to chiral system has been shown in previous studies in both the bulk \cite{Liebchen2017} and confined space \cite{caprini2019, caprini2021, snezhko2020, Yang2020}, and our simulations exhibit the same trend in confined systems. More recently, a work by Lei et al \cite{D2SM01402E} also studied the phase dependence of chiral active particles confined to a circular boundary. They assumed anisotropic interactions between the particles, instead of polar interactions considered in this study, and the interaction with the confinement wall in their case was completely repulsive in nature, and did not affect particle alignment. Strikingly however, they also observed phases analogous to edge currents, flocking and ordered oscillations found in this study, indicating that these phases are independent of interactions present in the system and the direct consequence of chirality.

One of the more notable observations in our simulations is that in the low \( \gamma_{p}, \gamma_{w} \) regime, the phase transition from edge currents to the flocking phase occurs at the same chirality, regardless of the particle number; in future work, by deriving a continuum hydrodynamic theory for this system, this phase transition behavior may be explained through a stability analysis. Furthermore, in a continuum model of chiral active fluid, viscous stress that does not result in dissipation (odd viscosity) but due to the reciprocal symmetry breaking is also involved in collective dynamics in bulk \cite{vitelli2017, yan2022, komura2021}. How such odd viscosity changes under boundary geometry remains a subject for future investigation.  

Even when the compartment inside living cells is symmetrical, there is often chirality in the cytoskeletal proteins and molecular motor proteins that show self-organized structures with active force generation \cite{Afroze2021}. Conversely, cytoskeletons beneath the cell membrane are also subject to precise regulation through protein interactions, resulting in context-dependent structures from a single set of proteins such as ring-like contractile gel and active retrograde flow \cite{heisenberg2021}. In circular cells, the interplay between the confining boundary and the chirality of the actin cytoskeleton can give rise to various patterns \cite{Tee2015}. This is also true at the multicellular scale for mammalian cells \cite{Wan2011}, and seen in edge currents for bacteria \cite{Beppu2021} and cell monolayers \cite{giomi2022}. Thus, manipulating the interaction between the inherent asymmetry of molecules such as chirality and boundary geometry may provide a deeper understanding of biological systems that generate emergent collective dynamics such as hydrodynamic bound states in swimming algae \cite{goldstein2009} and diverse ordered structures like living crystal forms \cite{libchaber2015, fakhri2022}. 

\section*{Acknowledgements}
This work was supported by Grant-in-Aid for Scientific Research on Innovative Areas 18H05427, Grant-in-Aid for Scientific Research (B) 20H01872, Grant-in-Aid for Challenging Research (Exploratory) from MEXT (to YTM), and JASSO Honors Scholarship (to AN).

\section*{Appendix}
\subsection{Quantitative definitions of the dynamical phases}
All the phases observed in this study show features of boundary flow, flocking, oscillation or a combination thereof. As such, it is possible to quantitatively define the dynamic steady states by utilizing multiple order parameters: vortex order parameter (VOP, \(\Phi_{vop}\)) and swarm order parameter (SOP, \(\psi_{sop}\)), as defined earlier and mean normalized tangential velocity \((v_T)\) as defined in the supplementary information \cite{supp}.

The phases in which all the particles move near the boundary, namely edge currents (counterclockwise and clockwise), and counterrotating currents, have a virtually constant \(\Phi_{vop}\), nearly equal to 1; therefore, the steady states with \( \langle \Phi_{vop} \rangle > 0.9 \) are either edge currents or counterrotating currents. To differentiate between these states, we use \(v_T\); counterclockwise edge currents have \( 0.9 < \langle v_T \rangle \leq 1 \), clockwise edge currents have \( -1 \leq \langle v_T \rangle < -0.9 \) and counterrotating currents have \( -0.5 < \langle v_T \rangle < 0 \). Note that for these phases, \( \langle \psi_{sop} \rangle \simeq 0 \).

The flocking phases can be easily defined by considering \(\psi_{sop}\); for the single flock phase, \( \langle \psi_{sop} \rangle > 0.9 \) and for the multiple flocks phase, \( 0.6 \leq \langle \psi_{sop} \rangle \leq 0.9 \).

The ordered oscillation and the counterrotation with oscillation phases can be characterized by the periodic oscillation of their VOP; the angular frequency of this oscillating \(\Phi_{vop}\), for both the phases is approximately twice the chirality, \( \omega_{\Phi_{vop}} = 2 \times \omega\), but for the ordered oscillation phase, \( \langle \Phi_{vop} \rangle < 0.2 \) whereas for the counterrotation with oscillation phase, \( 0.8 < \langle \Phi_{vop} \rangle < 0.9 \).

For defining the counterroatation with flocking, sustained spiral, travelling bands and disordered cluster phases, just a single parameter is not enough. In the counterroatation with flocking phase, \( 0.6 < \langle \Phi_{vop} \rangle < 0.9 \), and \( 0.3 < \langle \psi_{sop} \rangle < 0.6 \). For the sustained spiral phase,  \( 0.1 < \langle \Phi_{vop} \rangle < 0.6 \), and \( 0.2 < \langle \psi_{sop} \rangle < 0.4 \) and for the travelling bands phase, \( 0.3 < \langle \Phi_{vop} \rangle < 0.6 \), and \( 0.2 < \langle \psi_{sop} \rangle < 0.3 \) (see supplementary information \cite{supp}). To differentiate between the sustained spiral and the travelling bands phases, we again use \(v_T\); for the sustained spiral phase, while \( \langle v_T \rangle \) can take either positive or negative values, its has a significant magnitude, compared to the travelling bands phase, for which \( \langle v_T \rangle \simeq 0 \). For the disordered cluster phase, \( \langle \Phi_{vop} \rangle > 0.9 \), and \( 0.2 < \langle \psi_{sop} \rangle < 0.4 \) \cite{supp}.

With this, we have the quantitative definitions for all the ordered phases found in this study.

\subsection{Implementation of the random noise}
We have, \( \langle \eta_{m}(t) \eta_{n}(t') \rangle = 2 D \delta_{mn} \delta(t-t') \). From this, we obtain the noise as \(\eta_m = \sqrt{2 D dt} \times \xi \), where \(\xi\) is a pseudo-random number, drawn from a Gaussian distribution with mean 0 and standard deviation 1. The NumPy package of Python is used to generate the pseudo-random numbers, which utilizes the Mersenne Twister algorithm for this purpose \cite{harris2020array}.

% \nocite{*}
% \clearpage

\bibliography{main_ref}% Produces the bibliography via BibTeX.

\providecommand{\noopsort}[1]{}\providecommand{\singleletter}[1]{#1}%
\begin{thebibliography}{68}%
\makeatletter
\providecommand \@ifxundefined [1]{%
 \@ifx{#1\undefined}
}%
\providecommand \@ifnum [1]{%
 \ifnum #1\expandafter \@firstoftwo
 \else \expandafter \@secondoftwo
 \fi
}%
\providecommand \@ifx [1]{%
 \ifx #1\expandafter \@firstoftwo
 \else \expandafter \@secondoftwo
 \fi
}%
\providecommand \natexlab [1]{#1}%
\providecommand \enquote  [1]{``#1''}%
\providecommand \bibnamefont  [1]{#1}%
\providecommand \bibfnamefont [1]{#1}%
\providecommand \citenamefont [1]{#1}%
\providecommand \href@noop [0]{\@secondoftwo}%
\providecommand \href [0]{\begingroup \@sanitize@url \@href}%
\providecommand \@href[1]{\@@startlink{#1}\@@href}%
\providecommand \@@href[1]{\endgroup#1\@@endlink}%
\providecommand \@sanitize@url [0]{\catcode `\\12\catcode `\$12\catcode
  `\&12\catcode `\#12\catcode `\^12\catcode `\_12\catcode `\%12\relax}%
\providecommand \@@startlink[1]{}%
\providecommand \@@endlink[0]{}%
\providecommand \url  [0]{\begingroup\@sanitize@url \@url }%
\providecommand \@url [1]{\endgroup\@href {#1}{\urlprefix }}%
\providecommand \urlprefix  [0]{URL }%
\providecommand \Eprint [0]{\href }%
\providecommand \doibase [0]{https://doi.org/}%
\providecommand \selectlanguage [0]{\@gobble}%
\providecommand \bibinfo  [0]{\@secondoftwo}%
\providecommand \bibfield  [0]{\@secondoftwo}%
\providecommand \translation [1]{[#1]}%
\providecommand \BibitemOpen [0]{}%
\providecommand \bibitemStop [0]{}%
\providecommand \bibitemNoStop [0]{.\EOS\space}%
\providecommand \EOS [0]{\spacefactor3000\relax}%
\providecommand \BibitemShut  [1]{\csname bibitem#1\endcsname}%
\let\auto@bib@innerbib\@empty
%</preamble>
\bibitem [{\citenamefont {Ramaswamy}(2010)}]{ramaswamy2010}%
  \BibitemOpen
  \bibfield  {author} {\bibinfo {author} {\bibfnamefont {S.}~\bibnamefont
  {Ramaswamy}},\ }\bibfield  {title} {\bibinfo {title} {The mechanics and
  statistics of active matter},\ }\href
  {https://doi.org/10.1146/annurev-conmatphys-070909-104101} {\bibfield
  {journal} {\bibinfo  {journal} {Annu. Rev. Condens. Matter Phys.}\ }\textbf
  {\bibinfo {volume} {1}},\ \bibinfo {pages} {323} (\bibinfo {year}
  {2010})}\BibitemShut {NoStop}%
\bibitem [{\citenamefont {Marchetti}\ \emph {et~al.}(2013)\citenamefont
  {Marchetti}, \citenamefont {Joanny}, \citenamefont {Ramaswamy}, \citenamefont
  {Liverpool}, \citenamefont {Prost}, \citenamefont {Rao},\ and\ \citenamefont
  {Simha}}]{marchetti2013}%
  \BibitemOpen
  \bibfield  {author} {\bibinfo {author} {\bibfnamefont {M.~C.}\ \bibnamefont
  {Marchetti}}, \bibinfo {author} {\bibfnamefont {J.~F.}\ \bibnamefont
  {Joanny}}, \bibinfo {author} {\bibfnamefont {S.}~\bibnamefont {Ramaswamy}},
  \bibinfo {author} {\bibfnamefont {T.~B.}\ \bibnamefont {Liverpool}}, \bibinfo
  {author} {\bibfnamefont {J.}~\bibnamefont {Prost}}, \bibinfo {author}
  {\bibfnamefont {M.}~\bibnamefont {Rao}},\ and\ \bibinfo {author}
  {\bibfnamefont {R.~A.}\ \bibnamefont {Simha}},\ }\bibfield  {title} {\bibinfo
  {title} {Hydrodynamics of soft active matter},\ }\href
  {https://doi.org/10.1103/RevModPhys.85.1143} {\bibfield  {journal} {\bibinfo
  {journal} {Rev. Mod. Phys.}\ }\textbf {\bibinfo {volume} {85}},\ \bibinfo
  {pages} {1143} (\bibinfo {year} {2013})}\BibitemShut {NoStop}%
\bibitem [{\citenamefont {Shankar}\ \emph {et~al.}(2022)\citenamefont
  {Shankar}, \citenamefont {Souslov}, \citenamefont {Bowick}, \citenamefont
  {Marchetti},\ and\ \citenamefont {Vitelli}}]{shankar2022}%
  \BibitemOpen
  \bibfield  {author} {\bibinfo {author} {\bibfnamefont {S.}~\bibnamefont
  {Shankar}}, \bibinfo {author} {\bibfnamefont {A.}~\bibnamefont {Souslov}},
  \bibinfo {author} {\bibfnamefont {M.~J.}\ \bibnamefont {Bowick}}, \bibinfo
  {author} {\bibfnamefont {M.~C.}\ \bibnamefont {Marchetti}},\ and\ \bibinfo
  {author} {\bibfnamefont {V.}~\bibnamefont {Vitelli}},\ }\bibfield  {title}
  {\bibinfo {title} {Topological active matter},\ }\href
  {https://doi.org/10.1038/s42254-022-00445-3} {\bibfield  {journal} {\bibinfo
  {journal} {Nat. Rev. Phys.}\ }\textbf {\bibinfo {volume} {4}},\ \bibinfo
  {pages} {380} (\bibinfo {year} {2022})}\BibitemShut {NoStop}%
\bibitem [{\citenamefont {Ndlec}\ \emph {et~al.}(1997)\citenamefont {Ndlec},
  \citenamefont {Surrey}, \citenamefont {Maggs},\ and\ \citenamefont
  {Leibler}}]{leibler1997}%
  \BibitemOpen
  \bibfield  {author} {\bibinfo {author} {\bibfnamefont {F.~J.}\ \bibnamefont
  {Ndlec}}, \bibinfo {author} {\bibfnamefont {T.}~\bibnamefont {Surrey}},
  \bibinfo {author} {\bibfnamefont {A.~C.}\ \bibnamefont {Maggs}},\ and\
  \bibinfo {author} {\bibfnamefont {S.}~\bibnamefont {Leibler}},\ }\bibfield
  {title} {\bibinfo {title} {Self-organization of microtubules and motors},\
  }\href {https://doi.org/10.1038/38532} {\bibfield  {journal} {\bibinfo
  {journal} {Nature}\ }\textbf {\bibinfo {volume} {389}},\ \bibinfo {pages}
  {305} (\bibinfo {year} {1997})}\BibitemShut {NoStop}%
\bibitem [{\citenamefont {Schaller}\ \emph {et~al.}(2010)\citenamefont
  {Schaller}, \citenamefont {Weber}, \citenamefont {Semmrich}, \citenamefont
  {Frey},\ and\ \citenamefont {Bausch}}]{bausch2010}%
  \BibitemOpen
  \bibfield  {author} {\bibinfo {author} {\bibfnamefont {V.}~\bibnamefont
  {Schaller}}, \bibinfo {author} {\bibfnamefont {C.}~\bibnamefont {Weber}},
  \bibinfo {author} {\bibfnamefont {C.}~\bibnamefont {Semmrich}}, \bibinfo
  {author} {\bibfnamefont {E.}~\bibnamefont {Frey}},\ and\ \bibinfo {author}
  {\bibfnamefont {A.~R.}\ \bibnamefont {Bausch}},\ }\bibfield  {title}
  {\bibinfo {title} {Polar patterns of driven filaments},\ }\href
  {https://doi.org/10.1038/nature09312} {\bibfield  {journal} {\bibinfo
  {journal} {Nature}\ }\textbf {\bibinfo {volume} {467}},\ \bibinfo {pages}
  {73} (\bibinfo {year} {2010})}\BibitemShut {NoStop}%
\bibitem [{\citenamefont {Sanchez}\ \emph {et~al.}(2012)\citenamefont
  {Sanchez}, \citenamefont {Chen}, \citenamefont {DeCamp}, \citenamefont
  {Heymann},\ and\ \citenamefont {Dogic}}]{dogic2012}%
  \BibitemOpen
  \bibfield  {author} {\bibinfo {author} {\bibfnamefont {T.}~\bibnamefont
  {Sanchez}}, \bibinfo {author} {\bibfnamefont {D.~T.~N.}\ \bibnamefont
  {Chen}}, \bibinfo {author} {\bibfnamefont {S.~J.}\ \bibnamefont {DeCamp}},
  \bibinfo {author} {\bibfnamefont {M.}~\bibnamefont {Heymann}},\ and\ \bibinfo
  {author} {\bibfnamefont {Z.}~\bibnamefont {Dogic}},\ }\bibfield  {title}
  {\bibinfo {title} {Spontaneous motion in hierarchically assembled active
  matter},\ }\href {https://doi.org/10.1038/nature11591} {\bibfield  {journal}
  {\bibinfo  {journal} {Nature}\ }\textbf {\bibinfo {volume} {491}},\ \bibinfo
  {pages} {431} (\bibinfo {year} {2012})}\BibitemShut {NoStop}%
\bibitem [{\citenamefont {Aranson}(2022)}]{aronson2022}%
  \BibitemOpen
  \bibfield  {author} {\bibinfo {author} {\bibfnamefont {I.~S.}\ \bibnamefont
  {Aranson}},\ }\bibfield  {title} {\bibinfo {title} {Bacterial active
  matter},\ }\href {https://doi.org/10.1088/1361-6633/ac723d} {\bibfield
  {journal} {\bibinfo  {journal} {Rep. Prog. Phys.}\ }\textbf {\bibinfo
  {volume} {85}},\ \bibinfo {pages} {076601} (\bibinfo {year}
  {2022})}\BibitemShut {NoStop}%
\bibitem [{\citenamefont {Doostmohammadi}\ \emph {et~al.}(2018)\citenamefont
  {Doostmohammadi}, \citenamefont {Ign{\'{e}}s-Mullol}, \citenamefont
  {Yeomans},\ and\ \citenamefont {Sagu{\'{e}}s}}]{amin2018}%
  \BibitemOpen
  \bibfield  {author} {\bibinfo {author} {\bibfnamefont {A.}~\bibnamefont
  {Doostmohammadi}}, \bibinfo {author} {\bibfnamefont {J.}~\bibnamefont
  {Ign{\'{e}}s-Mullol}}, \bibinfo {author} {\bibfnamefont {J.~M.}\ \bibnamefont
  {Yeomans}},\ and\ \bibinfo {author} {\bibfnamefont {F.}~\bibnamefont
  {Sagu{\'{e}}s}},\ }\bibfield  {title} {\bibinfo {title} {Active nematics},\
  }\bibfield  {journal} {\bibinfo  {journal} {Nat. Commun.}\ }\textbf {\bibinfo
  {volume} {9}},\ \href {https://doi.org/10.1038/s41467-018-05666-8}
  {10.1038/s41467-018-05666-8} (\bibinfo {year} {2018})\BibitemShut {NoStop}%
\bibitem [{\citenamefont {Jhawar}\ \emph {et~al.}(2020)\citenamefont {Jhawar},
  \citenamefont {Morris}, \citenamefont {Amith-Kumar}, \citenamefont {Raj},
  \citenamefont {Rogers}, \citenamefont {Rajendran},\ and\ \citenamefont
  {Guttal}}]{guttal2020}%
  \BibitemOpen
  \bibfield  {author} {\bibinfo {author} {\bibfnamefont {J.}~\bibnamefont
  {Jhawar}}, \bibinfo {author} {\bibfnamefont {R.~G.}\ \bibnamefont {Morris}},
  \bibinfo {author} {\bibfnamefont {U.~R.}\ \bibnamefont {Amith-Kumar}},
  \bibinfo {author} {\bibfnamefont {M.~D.}\ \bibnamefont {Raj}}, \bibinfo
  {author} {\bibfnamefont {T.}~\bibnamefont {Rogers}}, \bibinfo {author}
  {\bibfnamefont {H.}~\bibnamefont {Rajendran}},\ and\ \bibinfo {author}
  {\bibfnamefont {V.}~\bibnamefont {Guttal}},\ }\bibfield  {title} {\bibinfo
  {title} {Noise-induced schooling of fish},\ }\href
  {https://doi.org/10.1038/s41567-020-0787-y} {\bibfield  {journal} {\bibinfo
  {journal} {Nat. Phys.}\ }\textbf {\bibinfo {volume} {16}},\ \bibinfo {pages}
  {488} (\bibinfo {year} {2020})}\BibitemShut {NoStop}%
\bibitem [{\citenamefont {Cavagna}\ and\ \citenamefont
  {Giardina}(2014)}]{giardina2014}%
  \BibitemOpen
  \bibfield  {author} {\bibinfo {author} {\bibfnamefont {A.}~\bibnamefont
  {Cavagna}}\ and\ \bibinfo {author} {\bibfnamefont {I.}~\bibnamefont
  {Giardina}},\ }\bibfield  {title} {\bibinfo {title} {Bird flocks as condensed
  matter},\ }\href {https://doi.org/10.1146/annurev-conmatphys-031113-133834}
  {\bibfield  {journal} {\bibinfo  {journal} {Annu. Rev. Condens. Matter
  Phys.}\ }\textbf {\bibinfo {volume} {5}},\ \bibinfo {pages} {183} (\bibinfo
  {year} {2014})}\BibitemShut {NoStop}%
\bibitem [{\citenamefont {Bain}\ and\ \citenamefont
  {Bartolo}(2019)}]{bartolo2019}%
  \BibitemOpen
  \bibfield  {author} {\bibinfo {author} {\bibfnamefont {N.}~\bibnamefont
  {Bain}}\ and\ \bibinfo {author} {\bibfnamefont {D.}~\bibnamefont {Bartolo}},\
  }\bibfield  {title} {\bibinfo {title} {Dynamic response and hydrodynamics of
  polarized crowds},\ }\href {https://doi.org/10.1126/science.aat9891}
  {\bibfield  {journal} {\bibinfo  {journal} {Science}\ }\textbf {\bibinfo
  {volume} {363}},\ \bibinfo {pages} {46} (\bibinfo {year} {2019})}\BibitemShut
  {NoStop}%
\bibitem [{\citenamefont {Vizsnyiczai}\ \emph {et~al.}(2017)\citenamefont
  {Vizsnyiczai}, \citenamefont {Frangipane}, \citenamefont {Maggi},
  \citenamefont {Saglimbeni}, \citenamefont {Bianchi},\ and\ \citenamefont
  {Leonardo}}]{leonardo2017}%
  \BibitemOpen
  \bibfield  {author} {\bibinfo {author} {\bibfnamefont {G.}~\bibnamefont
  {Vizsnyiczai}}, \bibinfo {author} {\bibfnamefont {G.}~\bibnamefont
  {Frangipane}}, \bibinfo {author} {\bibfnamefont {C.}~\bibnamefont {Maggi}},
  \bibinfo {author} {\bibfnamefont {F.}~\bibnamefont {Saglimbeni}}, \bibinfo
  {author} {\bibfnamefont {S.}~\bibnamefont {Bianchi}},\ and\ \bibinfo {author}
  {\bibfnamefont {R.~D.}\ \bibnamefont {Leonardo}},\ }\bibfield  {title}
  {\bibinfo {title} {Light controlled 3d micromotors powered by bacteria},\
  }\bibfield  {journal} {\bibinfo  {journal} {Nat. Commun.}\ }\textbf {\bibinfo
  {volume} {8}},\ \href {https://doi.org/10.1038/ncomms15974}
  {10.1038/ncomms15974} (\bibinfo {year} {2017})\BibitemShut {NoStop}%
\bibitem [{\citenamefont {Ross}\ \emph {et~al.}(2019)\citenamefont {Ross},
  \citenamefont {Lee}, \citenamefont {Qu}, \citenamefont {Banks}, \citenamefont
  {Phillips},\ and\ \citenamefont {Thomson}}]{thomson2019}%
  \BibitemOpen
  \bibfield  {author} {\bibinfo {author} {\bibfnamefont {T.~D.}\ \bibnamefont
  {Ross}}, \bibinfo {author} {\bibfnamefont {H.~J.}\ \bibnamefont {Lee}},
  \bibinfo {author} {\bibfnamefont {Z.}~\bibnamefont {Qu}}, \bibinfo {author}
  {\bibfnamefont {R.~A.}\ \bibnamefont {Banks}}, \bibinfo {author}
  {\bibfnamefont {R.}~\bibnamefont {Phillips}},\ and\ \bibinfo {author}
  {\bibfnamefont {M.}~\bibnamefont {Thomson}},\ }\bibfield  {title} {\bibinfo
  {title} {Controlling organization and forces in active matter through
  optically defined boundaries},\ }\href
  {https://doi.org/10.1038/s41586-019-1447-1} {\bibfield  {journal} {\bibinfo
  {journal} {Nature}\ }\textbf {\bibinfo {volume} {572}},\ \bibinfo {pages}
  {224} (\bibinfo {year} {2019})}\BibitemShut {NoStop}%
\bibitem [{\citenamefont {Woodhouse}\ and\ \citenamefont
  {Goldstein}(2012)}]{goldstein2012}%
  \BibitemOpen
  \bibfield  {author} {\bibinfo {author} {\bibfnamefont {F.~G.}\ \bibnamefont
  {Woodhouse}}\ and\ \bibinfo {author} {\bibfnamefont {R.~E.}\ \bibnamefont
  {Goldstein}},\ }\bibfield  {title} {\bibinfo {title} {Spontaneous circulation
  of confined active suspensions},\ }\href
  {https://doi.org/10.1103/PhysRevLett.109.168105} {\bibfield  {journal}
  {\bibinfo  {journal} {Phys. Rev. Lett.}\ }\textbf {\bibinfo {volume} {109}},\
  \bibinfo {pages} {168105} (\bibinfo {year} {2012})}\BibitemShut {NoStop}%
\bibitem [{\citenamefont {Wioland}\ \emph
  {et~al.}(2013{\natexlab{a}})\citenamefont {Wioland}, \citenamefont
  {Woodhouse}, \citenamefont {Dunkel}, \citenamefont {Kessler},\ and\
  \citenamefont {Goldstein}}]{goldstein2013}%
  \BibitemOpen
  \bibfield  {author} {\bibinfo {author} {\bibfnamefont {H.}~\bibnamefont
  {Wioland}}, \bibinfo {author} {\bibfnamefont {F.~G.}\ \bibnamefont
  {Woodhouse}}, \bibinfo {author} {\bibfnamefont {J.}~\bibnamefont {Dunkel}},
  \bibinfo {author} {\bibfnamefont {J.~O.}\ \bibnamefont {Kessler}},\ and\
  \bibinfo {author} {\bibfnamefont {R.~E.}\ \bibnamefont {Goldstein}},\
  }\bibfield  {title} {\bibinfo {title} {Confinement stabilizes a bacterial
  suspension into a spiral vortex},\ }\href
  {https://doi.org/10.1103/PhysRevLett.110.268102} {\bibfield  {journal}
  {\bibinfo  {journal} {Phys. Rev. Lett.}\ }\textbf {\bibinfo {volume} {110}},\
  \bibinfo {pages} {268102} (\bibinfo {year} {2013}{\natexlab{a}})}\BibitemShut
  {NoStop}%
\bibitem [{\citenamefont {Lushi}\ \emph {et~al.}(2014)\citenamefont {Lushi},
  \citenamefont {Wioland},\ and\ \citenamefont {Goldstein}}]{goldstein2014}%
  \BibitemOpen
  \bibfield  {author} {\bibinfo {author} {\bibfnamefont {E.}~\bibnamefont
  {Lushi}}, \bibinfo {author} {\bibfnamefont {H.}~\bibnamefont {Wioland}},\
  and\ \bibinfo {author} {\bibfnamefont {R.~E.}\ \bibnamefont {Goldstein}},\
  }\bibfield  {title} {\bibinfo {title} {Fluid flows created by swimming
  bacteria drive self-organization in confined suspensions},\ }\href
  {https://doi.org/10.1073/pnas.1405698111} {\bibfield  {journal} {\bibinfo
  {journal} {Proc. Natl. Acad. Sci. U.S.A.}\ }\textbf {\bibinfo {volume}
  {111}},\ \bibinfo {pages} {9733} (\bibinfo {year} {2014})}\BibitemShut
  {NoStop}%
\bibitem [{\citenamefont {Shendruk}\ \emph {et~al.}(2017)\citenamefont
  {Shendruk}, \citenamefont {Doostmohammadi}, \citenamefont {Thijssen},\ and\
  \citenamefont {Yeomans}}]{Shendruk2017}%
  \BibitemOpen
  \bibfield  {author} {\bibinfo {author} {\bibfnamefont {T.~N.}\ \bibnamefont
  {Shendruk}}, \bibinfo {author} {\bibfnamefont {A.}~\bibnamefont
  {Doostmohammadi}}, \bibinfo {author} {\bibfnamefont {K.}~\bibnamefont
  {Thijssen}},\ and\ \bibinfo {author} {\bibfnamefont {J.~M.}\ \bibnamefont
  {Yeomans}},\ }\bibfield  {title} {\bibinfo {title} {Dancing disclinations in
  confined active nematics},\ }\href {https://doi.org/10.1039/c6sm02310j}
  {\bibfield  {journal} {\bibinfo  {journal} {Soft Matter}\ }\textbf {\bibinfo
  {volume} {13}},\ \bibinfo {pages} {3853} (\bibinfo {year}
  {2017})}\BibitemShut {NoStop}%
\bibitem [{\citenamefont {Doostmohammadi}\ \emph {et~al.}(2017)\citenamefont
  {Doostmohammadi}, \citenamefont {Shendruk}, \citenamefont {Thijssen},\ and\
  \citenamefont {Yeomans}}]{Doostmohammadi2017}%
  \BibitemOpen
  \bibfield  {author} {\bibinfo {author} {\bibfnamefont {A.}~\bibnamefont
  {Doostmohammadi}}, \bibinfo {author} {\bibfnamefont {T.~N.}\ \bibnamefont
  {Shendruk}}, \bibinfo {author} {\bibfnamefont {K.}~\bibnamefont {Thijssen}},\
  and\ \bibinfo {author} {\bibfnamefont {J.~M.}\ \bibnamefont {Yeomans}},\
  }\bibfield  {title} {\bibinfo {title} {Onset of meso-scale turbulence in
  active nematics},\ }\bibfield  {journal} {\bibinfo  {journal} {Nat. Commun.}\
  }\textbf {\bibinfo {volume} {8}},\ \href
  {https://doi.org/10.1038/ncomms15326} {10.1038/ncomms15326} (\bibinfo {year}
  {2017})\BibitemShut {NoStop}%
\bibitem [{\citenamefont {Huang}\ \emph {et~al.}(2021)\citenamefont {Huang},
  \citenamefont {Du}, \citenamefont {Jiang},\ and\ \citenamefont
  {Hou}}]{Huang2021}%
  \BibitemOpen
  \bibfield  {author} {\bibinfo {author} {\bibfnamefont {D.}~\bibnamefont
  {Huang}}, \bibinfo {author} {\bibfnamefont {Y.}~\bibnamefont {Du}}, \bibinfo
  {author} {\bibfnamefont {H.}~\bibnamefont {Jiang}},\ and\ \bibinfo {author}
  {\bibfnamefont {Z.}~\bibnamefont {Hou}},\ }\bibfield  {title} {\bibinfo
  {title} {Emergent spiral vortex of confined biased active particles},\
  }\bibfield  {journal} {\bibinfo  {journal} {Phys. Rev. E}\ }\textbf {\bibinfo
  {volume} {104}},\ \href {https://doi.org/10.1103/physreve.104.034606}
  {10.1103/physreve.104.034606} (\bibinfo {year} {2021})\BibitemShut {NoStop}%
\bibitem [{\citenamefont {Wioland}\ \emph
  {et~al.}(2016{\natexlab{a}})\citenamefont {Wioland}, \citenamefont
  {Woodhouse}, \citenamefont {Dunkel},\ and\ \citenamefont
  {Goldstein}}]{goldstein2016a}%
  \BibitemOpen
  \bibfield  {author} {\bibinfo {author} {\bibfnamefont {H.}~\bibnamefont
  {Wioland}}, \bibinfo {author} {\bibfnamefont {F.~G.}\ \bibnamefont
  {Woodhouse}}, \bibinfo {author} {\bibfnamefont {J.}~\bibnamefont {Dunkel}},\
  and\ \bibinfo {author} {\bibfnamefont {R.~E.}\ \bibnamefont {Goldstein}},\
  }\bibfield  {title} {\bibinfo {title} {Ferromagnetic and antiferromagnetic
  order in bacterial vortex lattices},\ }\href
  {https://doi.org/10.1038/nphys3607} {\bibfield  {journal} {\bibinfo
  {journal} {Nat. Phys.}\ }\textbf {\bibinfo {volume} {12}},\ \bibinfo {pages}
  {341} (\bibinfo {year} {2016}{\natexlab{a}})}\BibitemShut {NoStop}%
\bibitem [{\citenamefont {Wioland}\ \emph
  {et~al.}(2016{\natexlab{b}})\citenamefont {Wioland}, \citenamefont {Lushi},\
  and\ \citenamefont {Goldstein}}]{goldstein2016b}%
  \BibitemOpen
  \bibfield  {author} {\bibinfo {author} {\bibfnamefont {H.}~\bibnamefont
  {Wioland}}, \bibinfo {author} {\bibfnamefont {E.}~\bibnamefont {Lushi}},\
  and\ \bibinfo {author} {\bibfnamefont {R.~E.}\ \bibnamefont {Goldstein}},\
  }\bibfield  {title} {\bibinfo {title} {Directed collective motion of bacteria
  under channel confinement},\ }\href
  {https://doi.org/10.1088/1367-2630/18/7/075002} {\bibfield  {journal}
  {\bibinfo  {journal} {New J. Phys.}\ }\textbf {\bibinfo {volume} {18}},\
  \bibinfo {pages} {075002} (\bibinfo {year} {2016}{\natexlab{b}})}\BibitemShut
  {NoStop}%
\bibitem [{\citenamefont {Beppu}\ \emph {et~al.}(2017)\citenamefont {Beppu},
  \citenamefont {Izri}, \citenamefont {Gohya}, \citenamefont {Eto},
  \citenamefont {Ichikawa},\ and\ \citenamefont {Maeda}}]{Beppu2017}%
  \BibitemOpen
  \bibfield  {author} {\bibinfo {author} {\bibfnamefont {K.}~\bibnamefont
  {Beppu}}, \bibinfo {author} {\bibfnamefont {Z.}~\bibnamefont {Izri}},
  \bibinfo {author} {\bibfnamefont {J.}~\bibnamefont {Gohya}}, \bibinfo
  {author} {\bibfnamefont {K.}~\bibnamefont {Eto}}, \bibinfo {author}
  {\bibfnamefont {M.}~\bibnamefont {Ichikawa}},\ and\ \bibinfo {author}
  {\bibfnamefont {Y.~T.}\ \bibnamefont {Maeda}},\ }\bibfield  {title} {\bibinfo
  {title} {Geometry-driven collective ordering of bacterial vortices},\ }\href
  {https://doi.org/10.1039/c7sm00999b} {\bibfield  {journal} {\bibinfo
  {journal} {Soft Matter}\ }\textbf {\bibinfo {volume} {13}},\ \bibinfo {pages}
  {5038} (\bibinfo {year} {2017})}\BibitemShut {NoStop}%
\bibitem [{\citenamefont {Nishiguchi}\ \emph {et~al.}(2018)\citenamefont
  {Nishiguchi}, \citenamefont {Aranson}, \citenamefont {Snezhko},\ and\
  \citenamefont {Sokolov}}]{nishiguchi2018}%
  \BibitemOpen
  \bibfield  {author} {\bibinfo {author} {\bibfnamefont {D.}~\bibnamefont
  {Nishiguchi}}, \bibinfo {author} {\bibfnamefont {I.~S.}\ \bibnamefont
  {Aranson}}, \bibinfo {author} {\bibfnamefont {A.}~\bibnamefont {Snezhko}},\
  and\ \bibinfo {author} {\bibfnamefont {A.}~\bibnamefont {Sokolov}},\
  }\bibfield  {title} {\bibinfo {title} {Engineering bacterial vortex lattice
  via direct laser lithography},\ }\bibfield  {journal} {\bibinfo  {journal}
  {Nat. Commun.}\ }\textbf {\bibinfo {volume} {9}},\ \href
  {https://doi.org/10.1038/s41467-018-06842-6} {10.1038/s41467-018-06842-6}
  (\bibinfo {year} {2018})\BibitemShut {NoStop}%
\bibitem [{\citenamefont {Reinken}\ \emph {et~al.}(2020)\citenamefont
  {Reinken}, \citenamefont {Nishiguchi}, \citenamefont {Heidenreich},
  \citenamefont {Sokolov}, \citenamefont {B\"{a}r}, \citenamefont {Klapp},\
  and\ \citenamefont {Aranson}}]{aronson2020}%
  \BibitemOpen
  \bibfield  {author} {\bibinfo {author} {\bibfnamefont {H.}~\bibnamefont
  {Reinken}}, \bibinfo {author} {\bibfnamefont {D.}~\bibnamefont {Nishiguchi}},
  \bibinfo {author} {\bibfnamefont {S.}~\bibnamefont {Heidenreich}}, \bibinfo
  {author} {\bibfnamefont {A.}~\bibnamefont {Sokolov}}, \bibinfo {author}
  {\bibfnamefont {M.}~\bibnamefont {B\"{a}r}}, \bibinfo {author} {\bibfnamefont
  {S.~H.~L.}\ \bibnamefont {Klapp}},\ and\ \bibinfo {author} {\bibfnamefont
  {I.~S.}\ \bibnamefont {Aranson}},\ }\bibfield  {title} {\bibinfo {title}
  {Organizing bacterial vortex lattices by periodic obstacle arrays},\
  }\bibfield  {journal} {\bibinfo  {journal} {Commun. Phys.}\ }\textbf
  {\bibinfo {volume} {3}},\ \href {https://doi.org/10.1038/s42005-020-0337-z}
  {10.1038/s42005-020-0337-z} (\bibinfo {year} {2020})\BibitemShut {NoStop}%
\bibitem [{\citenamefont {Beppu}\ \emph {et~al.}(2021)\citenamefont {Beppu},
  \citenamefont {Izri}, \citenamefont {Sato}, \citenamefont {Yamanishi},
  \citenamefont {Sumino},\ and\ \citenamefont {Maeda}}]{Beppu2021}%
  \BibitemOpen
  \bibfield  {author} {\bibinfo {author} {\bibfnamefont {K.}~\bibnamefont
  {Beppu}}, \bibinfo {author} {\bibfnamefont {Z.}~\bibnamefont {Izri}},
  \bibinfo {author} {\bibfnamefont {T.}~\bibnamefont {Sato}}, \bibinfo {author}
  {\bibfnamefont {Y.}~\bibnamefont {Yamanishi}}, \bibinfo {author}
  {\bibfnamefont {Y.}~\bibnamefont {Sumino}},\ and\ \bibinfo {author}
  {\bibfnamefont {Y.~T.}\ \bibnamefont {Maeda}},\ }\bibfield  {title} {\bibinfo
  {title} {Edge current and pairing order transition in chiral bacterial
  vortices},\ }\bibfield  {journal} {\bibinfo  {journal} {Proc. Natl. Acad.
  Sci. U.S.A.}\ }\textbf {\bibinfo {volume} {118}},\ \href
  {https://doi.org/10.1073/pnas.2107461118} {10.1073/pnas.2107461118} (\bibinfo
  {year} {2021})\BibitemShut {NoStop}%
\bibitem [{\citenamefont {Opathalage}\ \emph {et~al.}(2019)\citenamefont
  {Opathalage}, \citenamefont {Norton}, \citenamefont {Juniper}, \citenamefont
  {Langeslay}, \citenamefont {Aghvami}, \citenamefont {Fraden},\ and\
  \citenamefont {Dogic}}]{Opathalage2019}%
  \BibitemOpen
  \bibfield  {author} {\bibinfo {author} {\bibfnamefont {A.}~\bibnamefont
  {Opathalage}}, \bibinfo {author} {\bibfnamefont {M.~M.}\ \bibnamefont
  {Norton}}, \bibinfo {author} {\bibfnamefont {M.~P.~N.}\ \bibnamefont
  {Juniper}}, \bibinfo {author} {\bibfnamefont {B.}~\bibnamefont {Langeslay}},
  \bibinfo {author} {\bibfnamefont {S.~A.}\ \bibnamefont {Aghvami}}, \bibinfo
  {author} {\bibfnamefont {S.}~\bibnamefont {Fraden}},\ and\ \bibinfo {author}
  {\bibfnamefont {Z.}~\bibnamefont {Dogic}},\ }\bibfield  {title} {\bibinfo
  {title} {Self-organized dynamics and the transition to turbulence of confined
  active nematics},\ }\href {https://doi.org/10.1073/pnas.1816733116}
  {\bibfield  {journal} {\bibinfo  {journal} {Proc. Natl. Acad. Sci. U.S.A.}\
  }\textbf {\bibinfo {volume} {116}},\ \bibinfo {pages} {4788} (\bibinfo {year}
  {2019})}\BibitemShut {NoStop}%
\bibitem [{\citenamefont {Guillamat}\ \emph {et~al.}(2017)\citenamefont
  {Guillamat}, \citenamefont {Ign{\'{e}}s-Mullol},\ and\ \citenamefont
  {Sagu{\'{e}}s}}]{Guillamat2017}%
  \BibitemOpen
  \bibfield  {author} {\bibinfo {author} {\bibfnamefont {P.}~\bibnamefont
  {Guillamat}}, \bibinfo {author} {\bibfnamefont {J.}~\bibnamefont
  {Ign{\'{e}}s-Mullol}},\ and\ \bibinfo {author} {\bibfnamefont
  {F.}~\bibnamefont {Sagu{\'{e}}s}},\ }\bibfield  {title} {\bibinfo {title}
  {Taming active turbulence with patterned soft interfaces},\ }\bibfield
  {journal} {\bibinfo  {journal} {Nat. Commun.}\ }\textbf {\bibinfo {volume}
  {8}},\ \href {https://doi.org/10.1038/s41467-017-00617-1}
  {10.1038/s41467-017-00617-1} (\bibinfo {year} {2017})\BibitemShut {NoStop}%
\bibitem [{\citenamefont {Araki}\ \emph {et~al.}(2021)\citenamefont {Araki},
  \citenamefont {Beppu}, \citenamefont {Kabir}, \citenamefont {Kakugo},\ and\
  \citenamefont {Maeda}}]{araki2021}%
  \BibitemOpen
  \bibfield  {author} {\bibinfo {author} {\bibfnamefont {S.}~\bibnamefont
  {Araki}}, \bibinfo {author} {\bibfnamefont {K.}~\bibnamefont {Beppu}},
  \bibinfo {author} {\bibfnamefont {A.~M.~R.}\ \bibnamefont {Kabir}}, \bibinfo
  {author} {\bibfnamefont {A.}~\bibnamefont {Kakugo}},\ and\ \bibinfo {author}
  {\bibfnamefont {Y.~T.}\ \bibnamefont {Maeda}},\ }\bibfield  {title} {\bibinfo
  {title} {Controlling collective motion of kinesin-driven microtubules via
  patterning of topographic landscapes},\ }\href
  {https://doi.org/10.1021/acs.nanolett.1c03952} {\bibfield  {journal}
  {\bibinfo  {journal} {Nano Lett.}\ }\textbf {\bibinfo {volume} {21}},\
  \bibinfo {pages} {10478} (\bibinfo {year} {2021})}\BibitemShut {NoStop}%
\bibitem [{\citenamefont {Doxzen}\ \emph {et~al.}(2013)\citenamefont {Doxzen},
  \citenamefont {Vedula}, \citenamefont {Leong}, \citenamefont {Hirata},
  \citenamefont {Gov}, \citenamefont {Kabla}, \citenamefont {Ladoux},\ and\
  \citenamefont {Lim}}]{Doxzen2013}%
  \BibitemOpen
  \bibfield  {author} {\bibinfo {author} {\bibfnamefont {K.}~\bibnamefont
  {Doxzen}}, \bibinfo {author} {\bibfnamefont {S.~R.~K.}\ \bibnamefont
  {Vedula}}, \bibinfo {author} {\bibfnamefont {M.~C.}\ \bibnamefont {Leong}},
  \bibinfo {author} {\bibfnamefont {H.}~\bibnamefont {Hirata}}, \bibinfo
  {author} {\bibfnamefont {N.~S.}\ \bibnamefont {Gov}}, \bibinfo {author}
  {\bibfnamefont {A.~J.}\ \bibnamefont {Kabla}}, \bibinfo {author}
  {\bibfnamefont {B.}~\bibnamefont {Ladoux}},\ and\ \bibinfo {author}
  {\bibfnamefont {C.~T.}\ \bibnamefont {Lim}},\ }\bibfield  {title} {\bibinfo
  {title} {Guidance of collective cell migration by substrate geometry},\
  }\href {https://doi.org/10.1039/c3ib40054a} {\bibfield  {journal} {\bibinfo
  {journal} {Integr. Biol.}\ }\textbf {\bibinfo {volume} {5}},\ \bibinfo
  {pages} {1026} (\bibinfo {year} {2013})}\BibitemShut {NoStop}%
\bibitem [{\citenamefont {DeCamp}\ \emph {et~al.}(2015)\citenamefont {DeCamp},
  \citenamefont {Redner}, \citenamefont {Baskaran}, \citenamefont {Hagan},\
  and\ \citenamefont {Dogic}}]{DeCamp2015}%
  \BibitemOpen
  \bibfield  {author} {\bibinfo {author} {\bibfnamefont {S.~J.}\ \bibnamefont
  {DeCamp}}, \bibinfo {author} {\bibfnamefont {G.~S.}\ \bibnamefont {Redner}},
  \bibinfo {author} {\bibfnamefont {A.}~\bibnamefont {Baskaran}}, \bibinfo
  {author} {\bibfnamefont {M.~F.}\ \bibnamefont {Hagan}},\ and\ \bibinfo
  {author} {\bibfnamefont {Z.}~\bibnamefont {Dogic}},\ }\bibfield  {title}
  {\bibinfo {title} {Orientational order of motile defects in
  active~nematics},\ }\href {https://doi.org/10.1038/nmat4387} {\bibfield
  {journal} {\bibinfo  {journal} {Nat. Mater.}\ }\textbf {\bibinfo {volume}
  {14}},\ \bibinfo {pages} {1110} (\bibinfo {year} {2015})}\BibitemShut
  {NoStop}%
\bibitem [{\citenamefont {Duclos}\ \emph {et~al.}(2018)\citenamefont {Duclos},
  \citenamefont {Blanch-Mercader}, \citenamefont {Yashunsky}, \citenamefont
  {Salbreux}, \citenamefont {Joanny}, \citenamefont {Prost},\ and\
  \citenamefont {Silberzan}}]{silberzan2018}%
  \BibitemOpen
  \bibfield  {author} {\bibinfo {author} {\bibfnamefont {G.}~\bibnamefont
  {Duclos}}, \bibinfo {author} {\bibfnamefont {C.}~\bibnamefont
  {Blanch-Mercader}}, \bibinfo {author} {\bibfnamefont {V.}~\bibnamefont
  {Yashunsky}}, \bibinfo {author} {\bibfnamefont {G.}~\bibnamefont {Salbreux}},
  \bibinfo {author} {\bibfnamefont {J.-F.}\ \bibnamefont {Joanny}}, \bibinfo
  {author} {\bibfnamefont {J.}~\bibnamefont {Prost}},\ and\ \bibinfo {author}
  {\bibfnamefont {P.}~\bibnamefont {Silberzan}},\ }\bibfield  {title} {\bibinfo
  {title} {Spontaneous shear flow in confined cellular nematics},\ }\href
  {https://doi.org/10.1038/s41567-018-0099-7} {\bibfield  {journal} {\bibinfo
  {journal} {Nat. Phys.}\ }\textbf {\bibinfo {volume} {14}},\ \bibinfo {pages}
  {728} (\bibinfo {year} {2018})}\BibitemShut {NoStop}%
\bibitem [{\citenamefont {Guillamat}\ \emph {et~al.}(2022)\citenamefont
  {Guillamat}, \citenamefont {Blanch-Mercader}, \citenamefont {Pernollet},
  \citenamefont {Kruse},\ and\ \citenamefont {Roux}}]{roux2022}%
  \BibitemOpen
  \bibfield  {author} {\bibinfo {author} {\bibfnamefont {P.}~\bibnamefont
  {Guillamat}}, \bibinfo {author} {\bibfnamefont {C.}~\bibnamefont
  {Blanch-Mercader}}, \bibinfo {author} {\bibfnamefont {G.}~\bibnamefont
  {Pernollet}}, \bibinfo {author} {\bibfnamefont {K.}~\bibnamefont {Kruse}},\
  and\ \bibinfo {author} {\bibfnamefont {A.}~\bibnamefont {Roux}},\ }\bibfield
  {title} {\bibinfo {title} {Integer topological defects organize stresses
  driving tissue morphogenesis},\ }\href
  {https://doi.org/10.1038/s41563-022-01194-5} {\bibfield  {journal} {\bibinfo
  {journal} {Nat. Mater.}\ }\textbf {\bibinfo {volume} {21}},\ \bibinfo {pages}
  {588} (\bibinfo {year} {2022})}\BibitemShut {NoStop}%
\bibitem [{\citenamefont {Keber}\ \emph {et~al.}(2014)\citenamefont {Keber},
  \citenamefont {Loiseau}, \citenamefont {Sanchez}, \citenamefont {DeCamp},
  \citenamefont {Giomi}, \citenamefont {Bowick}, \citenamefont {Marchetti},
  \citenamefont {Dogic},\ and\ \citenamefont {Bausch}}]{Keber2014}%
  \BibitemOpen
  \bibfield  {author} {\bibinfo {author} {\bibfnamefont {F.~C.}\ \bibnamefont
  {Keber}}, \bibinfo {author} {\bibfnamefont {E.}~\bibnamefont {Loiseau}},
  \bibinfo {author} {\bibfnamefont {T.}~\bibnamefont {Sanchez}}, \bibinfo
  {author} {\bibfnamefont {S.~J.}\ \bibnamefont {DeCamp}}, \bibinfo {author}
  {\bibfnamefont {L.}~\bibnamefont {Giomi}}, \bibinfo {author} {\bibfnamefont
  {M.~J.}\ \bibnamefont {Bowick}}, \bibinfo {author} {\bibfnamefont {M.~C.}\
  \bibnamefont {Marchetti}}, \bibinfo {author} {\bibfnamefont {Z.}~\bibnamefont
  {Dogic}},\ and\ \bibinfo {author} {\bibfnamefont {A.~R.}\ \bibnamefont
  {Bausch}},\ }\bibfield  {title} {\bibinfo {title} {Topology and dynamics of
  active nematic vesicles},\ }\href {https://doi.org/10.1126/science.1254784}
  {\bibfield  {journal} {\bibinfo  {journal} {Science}\ }\textbf {\bibinfo
  {volume} {345}},\ \bibinfo {pages} {1135} (\bibinfo {year}
  {2014})}\BibitemShut {NoStop}%
\bibitem [{\citenamefont {Hsu}\ \emph {et~al.}(2022)\citenamefont {Hsu},
  \citenamefont {Sciortino}, \citenamefont {de~la Trobe},\ and\ \citenamefont
  {Bausch}}]{Hsu2022}%
  \BibitemOpen
  \bibfield  {author} {\bibinfo {author} {\bibfnamefont {C.-P.}\ \bibnamefont
  {Hsu}}, \bibinfo {author} {\bibfnamefont {A.}~\bibnamefont {Sciortino}},
  \bibinfo {author} {\bibfnamefont {Y.~A.}\ \bibnamefont {de~la Trobe}},\ and\
  \bibinfo {author} {\bibfnamefont {A.~R.}\ \bibnamefont {Bausch}},\ }\bibfield
   {title} {\bibinfo {title} {Activity-induced polar patterns of filaments
  gliding on a sphere},\ }\bibfield  {journal} {\bibinfo  {journal} {Nat.
  Commun.}\ }\textbf {\bibinfo {volume} {13}},\ \href
  {https://doi.org/10.1038/s41467-022-30128-7} {10.1038/s41467-022-30128-7}
  (\bibinfo {year} {2022})\BibitemShut {NoStop}%
\bibitem [{\citenamefont {Liebchen}\ and\ \citenamefont
  {Levis}(2022)}]{libchen2022}%
  \BibitemOpen
  \bibfield  {author} {\bibinfo {author} {\bibfnamefont {B.}~\bibnamefont
  {Liebchen}}\ and\ \bibinfo {author} {\bibfnamefont {D.}~\bibnamefont
  {Levis}},\ }\bibfield  {title} {\bibinfo {title} {Chiral active matter},\
  }\href {https://doi.org/10.1209/0295-5075/ac8f69} {\bibfield  {journal}
  {\bibinfo  {journal} {EPL}\ }\textbf {\bibinfo {volume} {139}},\ \bibinfo
  {pages} {67001} (\bibinfo {year} {2022})}\BibitemShut {NoStop}%
\bibitem [{\citenamefont {Caprini}\ and\ \citenamefont
  {Marconi}(2019)}]{caprini2019}%
  \BibitemOpen
  \bibfield  {author} {\bibinfo {author} {\bibfnamefont {L.}~\bibnamefont
  {Caprini}}\ and\ \bibinfo {author} {\bibfnamefont {U.~M.~B.}\ \bibnamefont
  {Marconi}},\ }\bibfield  {title} {\bibinfo {title} {Active chiral particles
  under confinement: surface currents and bulk accumulation phenomena},\ }\href
  {https://doi.org/10.1039/c8sm02492h} {\bibfield  {journal} {\bibinfo
  {journal} {Soft Matter}\ }\textbf {\bibinfo {volume} {15}},\ \bibinfo {pages}
  {2627} (\bibinfo {year} {2019})}\BibitemShut {NoStop}%
\bibitem [{\citenamefont {Caprini}\ \emph {et~al.}(2021)\citenamefont
  {Caprini}, \citenamefont {Maggi},\ and\ \citenamefont
  {Marconi}}]{caprini2021}%
  \BibitemOpen
  \bibfield  {author} {\bibinfo {author} {\bibfnamefont {L.}~\bibnamefont
  {Caprini}}, \bibinfo {author} {\bibfnamefont {C.}~\bibnamefont {Maggi}},\
  and\ \bibinfo {author} {\bibfnamefont {U.~M.~B.}\ \bibnamefont {Marconi}},\
  }\bibfield  {title} {\bibinfo {title} {Collective effects in confined active
  brownian particles},\ }\href {https://doi.org/10.1063/5.0051315} {\bibfield
  {journal} {\bibinfo  {journal} {J. Chem. Phys.}\ }\textbf {\bibinfo {volume}
  {154}},\ \bibinfo {pages} {244901} (\bibinfo {year} {2021})}\BibitemShut
  {NoStop}%
\bibitem [{\citenamefont {Zhang}\ \emph
  {et~al.}(2020{\natexlab{a}})\citenamefont {Zhang}, \citenamefont {Hilton},
  \citenamefont {Short}, \citenamefont {Souslov},\ and\ \citenamefont
  {Snezhko}}]{snezhko2020}%
  \BibitemOpen
  \bibfield  {author} {\bibinfo {author} {\bibfnamefont {B.}~\bibnamefont
  {Zhang}}, \bibinfo {author} {\bibfnamefont {B.}~\bibnamefont {Hilton}},
  \bibinfo {author} {\bibfnamefont {C.}~\bibnamefont {Short}}, \bibinfo
  {author} {\bibfnamefont {A.}~\bibnamefont {Souslov}},\ and\ \bibinfo {author}
  {\bibfnamefont {A.}~\bibnamefont {Snezhko}},\ }\bibfield  {title} {\bibinfo
  {title} {Oscillatory chiral flows in confined active fluids with obstacles},\
  }\href {https://doi.org/10.1103/PhysRevResearch.2.043225} {\bibfield
  {journal} {\bibinfo  {journal} {Phys. Rev. Res.}\ }\textbf {\bibinfo {volume}
  {2}},\ \bibinfo {pages} {043225} (\bibinfo {year}
  {2020}{\natexlab{a}})}\BibitemShut {NoStop}%
\bibitem [{\citenamefont {K\"{u}mmel}\ \emph {et~al.}(2013)\citenamefont
  {K\"{u}mmel}, \citenamefont {ten Hagen}, \citenamefont {Wittkowski},
  \citenamefont {Buttinoni}, \citenamefont {Eichhorn}, \citenamefont {Volpe},
  \citenamefont {L\"{o}wen},\ and\ \citenamefont {Bechinger}}]{Kmmel2013}%
  \BibitemOpen
  \bibfield  {author} {\bibinfo {author} {\bibfnamefont {F.}~\bibnamefont
  {K\"{u}mmel}}, \bibinfo {author} {\bibfnamefont {B.}~\bibnamefont {ten
  Hagen}}, \bibinfo {author} {\bibfnamefont {R.}~\bibnamefont {Wittkowski}},
  \bibinfo {author} {\bibfnamefont {I.}~\bibnamefont {Buttinoni}}, \bibinfo
  {author} {\bibfnamefont {R.}~\bibnamefont {Eichhorn}}, \bibinfo {author}
  {\bibfnamefont {G.}~\bibnamefont {Volpe}}, \bibinfo {author} {\bibfnamefont
  {H.}~\bibnamefont {L\"{o}wen}},\ and\ \bibinfo {author} {\bibfnamefont
  {C.}~\bibnamefont {Bechinger}},\ }\bibfield  {title} {\bibinfo {title}
  {Circular motion of asymmetric self-propelling particles},\ }\bibfield
  {journal} {\bibinfo  {journal} {Phys. Rev. Lett.}\ }\textbf {\bibinfo
  {volume} {110}},\ \href {https://doi.org/10.1103/physrevlett.110.198302}
  {10.1103/physrevlett.110.198302} (\bibinfo {year} {2013})\BibitemShut
  {NoStop}%
\bibitem [{\citenamefont {Zhang}\ \emph
  {et~al.}(2020{\natexlab{b}})\citenamefont {Zhang}, \citenamefont {Sokolov},\
  and\ \citenamefont {Snezhko}}]{Zhang2020}%
  \BibitemOpen
  \bibfield  {author} {\bibinfo {author} {\bibfnamefont {B.}~\bibnamefont
  {Zhang}}, \bibinfo {author} {\bibfnamefont {A.}~\bibnamefont {Sokolov}},\
  and\ \bibinfo {author} {\bibfnamefont {A.}~\bibnamefont {Snezhko}},\
  }\bibfield  {title} {\bibinfo {title} {Reconfigurable emergent patterns in
  active chiral fluids},\ }\bibfield  {journal} {\bibinfo  {journal} {Nat.
  Commun.}\ }\textbf {\bibinfo {volume} {11}},\ \href
  {https://doi.org/10.1038/s41467-020-18209-x} {10.1038/s41467-020-18209-x}
  (\bibinfo {year} {2020}{\natexlab{b}})\BibitemShut {NoStop}%
\bibitem [{\citenamefont {Scholz}\ \emph {et~al.}(2018)\citenamefont {Scholz},
  \citenamefont {Engel},\ and\ \citenamefont {P\"{o}schel}}]{Scholz2018}%
  \BibitemOpen
  \bibfield  {author} {\bibinfo {author} {\bibfnamefont {C.}~\bibnamefont
  {Scholz}}, \bibinfo {author} {\bibfnamefont {M.}~\bibnamefont {Engel}},\ and\
  \bibinfo {author} {\bibfnamefont {T.}~\bibnamefont {P\"{o}schel}},\
  }\bibfield  {title} {\bibinfo {title} {Rotating robots move collectively and
  self-organize},\ }\bibfield  {journal} {\bibinfo  {journal} {Nat. Commun.}\
  }\textbf {\bibinfo {volume} {9}},\ \href
  {https://doi.org/10.1038/s41467-018-03154-7} {10.1038/s41467-018-03154-7}
  (\bibinfo {year} {2018})\BibitemShut {NoStop}%
\bibitem [{\citenamefont {Yang}\ \emph {et~al.}(2020)\citenamefont {Yang},
  \citenamefont {Ren}, \citenamefont {Cheng},\ and\ \citenamefont
  {Zhang}}]{Yang2020}%
  \BibitemOpen
  \bibfield  {author} {\bibinfo {author} {\bibfnamefont {X.}~\bibnamefont
  {Yang}}, \bibinfo {author} {\bibfnamefont {C.}~\bibnamefont {Ren}}, \bibinfo
  {author} {\bibfnamefont {K.}~\bibnamefont {Cheng}},\ and\ \bibinfo {author}
  {\bibfnamefont {H.~P.}\ \bibnamefont {Zhang}},\ }\bibfield  {title} {\bibinfo
  {title} {Robust boundary flow in chiral active fluid},\ }\bibfield  {journal}
  {\bibinfo  {journal} {Phys. Rev. E}\ }\textbf {\bibinfo {volume} {101}},\
  \href {https://doi.org/10.1103/physreve.101.022603}
  {10.1103/physreve.101.022603} (\bibinfo {year} {2020})\BibitemShut {NoStop}%
\bibitem [{\citenamefont {Liu}\ \emph {et~al.}(2020)\citenamefont {Liu},
  \citenamefont {Zhu}, \citenamefont {Zeng}, \citenamefont {Du}, \citenamefont
  {Ning}, \citenamefont {Wang}, \citenamefont {Chen}, \citenamefont {Lu},
  \citenamefont {Zheng}, \citenamefont {Ye},\ and\ \citenamefont
  {Yang}}]{Liu2020}%
  \BibitemOpen
  \bibfield  {author} {\bibinfo {author} {\bibfnamefont {P.}~\bibnamefont
  {Liu}}, \bibinfo {author} {\bibfnamefont {H.}~\bibnamefont {Zhu}}, \bibinfo
  {author} {\bibfnamefont {Y.}~\bibnamefont {Zeng}}, \bibinfo {author}
  {\bibfnamefont {G.}~\bibnamefont {Du}}, \bibinfo {author} {\bibfnamefont
  {L.}~\bibnamefont {Ning}}, \bibinfo {author} {\bibfnamefont {D.}~\bibnamefont
  {Wang}}, \bibinfo {author} {\bibfnamefont {K.}~\bibnamefont {Chen}}, \bibinfo
  {author} {\bibfnamefont {Y.}~\bibnamefont {Lu}}, \bibinfo {author}
  {\bibfnamefont {N.}~\bibnamefont {Zheng}}, \bibinfo {author} {\bibfnamefont
  {F.}~\bibnamefont {Ye}},\ and\ \bibinfo {author} {\bibfnamefont
  {M.}~\bibnamefont {Yang}},\ }\bibfield  {title} {\bibinfo {title}
  {Oscillating collective motion of active rotors in confinement},\ }\href
  {https://doi.org/10.1073/pnas.1922633117} {\bibfield  {journal} {\bibinfo
  {journal} {Proc. Natl. Acad. Sci. U.S.A.}\ }\textbf {\bibinfo {volume}
  {117}},\ \bibinfo {pages} {11901} (\bibinfo {year} {2020})}\BibitemShut
  {NoStop}%
\bibitem [{\citenamefont {Afroze}\ \emph {et~al.}(2021)\citenamefont {Afroze},
  \citenamefont {Inoue}, \citenamefont {Farhana}, \citenamefont {Hiraiwa},
  \citenamefont {Akiyama}, \citenamefont {Kabir}, \citenamefont {Sada},\ and\
  \citenamefont {Kakugo}}]{Afroze2021}%
  \BibitemOpen
  \bibfield  {author} {\bibinfo {author} {\bibfnamefont {F.}~\bibnamefont
  {Afroze}}, \bibinfo {author} {\bibfnamefont {D.}~\bibnamefont {Inoue}},
  \bibinfo {author} {\bibfnamefont {T.~I.}\ \bibnamefont {Farhana}}, \bibinfo
  {author} {\bibfnamefont {T.}~\bibnamefont {Hiraiwa}}, \bibinfo {author}
  {\bibfnamefont {R.}~\bibnamefont {Akiyama}}, \bibinfo {author} {\bibfnamefont
  {A.~M.~R.}\ \bibnamefont {Kabir}}, \bibinfo {author} {\bibfnamefont
  {K.}~\bibnamefont {Sada}},\ and\ \bibinfo {author} {\bibfnamefont
  {A.}~\bibnamefont {Kakugo}},\ }\bibfield  {title} {\bibinfo {title}
  {Monopolar flocking of microtubules in collective motion},\ }\href
  {https://doi.org/10.1016/j.bbrc.2021.05.037} {\bibfield  {journal} {\bibinfo
  {journal} {Biochem. Biophys. Res. Commun.}\ }\textbf {\bibinfo {volume}
  {563}},\ \bibinfo {pages} {73} (\bibinfo {year} {2021})}\BibitemShut
  {NoStop}%
\bibitem [{\citenamefont {Liebchen}\ and\ \citenamefont
  {Levis}(2017)}]{Liebchen2017}%
  \BibitemOpen
  \bibfield  {author} {\bibinfo {author} {\bibfnamefont {B.}~\bibnamefont
  {Liebchen}}\ and\ \bibinfo {author} {\bibfnamefont {D.}~\bibnamefont
  {Levis}},\ }\bibfield  {title} {\bibinfo {title} {Collective behavior of
  chiral active matter: Pattern formation and enhanced flocking},\ }\bibfield
  {journal} {\bibinfo  {journal} {Phys. Rev. Lett.}\ }\textbf {\bibinfo
  {volume} {119}},\ \href {https://doi.org/10.1103/physrevlett.119.058002}
  {10.1103/physrevlett.119.058002} (\bibinfo {year} {2017})\BibitemShut
  {NoStop}%
\bibitem [{\citenamefont {Levis}\ and\ \citenamefont
  {Liebchen}(2019)}]{Levis2019a}%
  \BibitemOpen
  \bibfield  {author} {\bibinfo {author} {\bibfnamefont {D.}~\bibnamefont
  {Levis}}\ and\ \bibinfo {author} {\bibfnamefont {B.}~\bibnamefont
  {Liebchen}},\ }\bibfield  {title} {\bibinfo {title} {Simultaneous phase
  separation and pattern formation in chiral active mixtures},\ }\bibfield
  {journal} {\bibinfo  {journal} {Phys. Rev. E}\ }\textbf {\bibinfo {volume}
  {100}},\ \href {https://doi.org/10.1103/physreve.100.012406}
  {10.1103/physreve.100.012406} (\bibinfo {year} {2019})\BibitemShut {NoStop}%
\bibitem [{\citenamefont {Levis}\ \emph {et~al.}(2019)\citenamefont {Levis},
  \citenamefont {Pagonabarraga},\ and\ \citenamefont {Liebchen}}]{Levis2019b}%
  \BibitemOpen
  \bibfield  {author} {\bibinfo {author} {\bibfnamefont {D.}~\bibnamefont
  {Levis}}, \bibinfo {author} {\bibfnamefont {I.}~\bibnamefont
  {Pagonabarraga}},\ and\ \bibinfo {author} {\bibfnamefont {B.}~\bibnamefont
  {Liebchen}},\ }\bibfield  {title} {\bibinfo {title} {Activity induced
  synchronization: Mutual flocking and chiral self-sorting},\ }\bibfield
  {journal} {\bibinfo  {journal} {Phys. Rev. Res.}\ }\textbf {\bibinfo {volume}
  {1}},\ \href {https://doi.org/10.1103/physrevresearch.1.023026}
  {10.1103/physrevresearch.1.023026} (\bibinfo {year} {2019})\BibitemShut
  {NoStop}%
\bibitem [{\citenamefont {Liao}\ and\ \citenamefont {Klapp}(2021)}]{Liao2021}%
  \BibitemOpen
  \bibfield  {author} {\bibinfo {author} {\bibfnamefont {G.-J.}\ \bibnamefont
  {Liao}}\ and\ \bibinfo {author} {\bibfnamefont {S.~H.~L.}\ \bibnamefont
  {Klapp}},\ }\bibfield  {title} {\bibinfo {title} {Emergent vortices and phase
  separation in systems of chiral active particles with dipolar interactions},\
  }\href {https://doi.org/10.1039/d1sm00545f} {\bibfield  {journal} {\bibinfo
  {journal} {Soft Matter}\ }\textbf {\bibinfo {volume} {17}},\ \bibinfo {pages}
  {6833} (\bibinfo {year} {2021})}\BibitemShut {NoStop}%
\bibitem [{\citenamefont {Ventejou}\ \emph {et~al.}(2021)\citenamefont
  {Ventejou}, \citenamefont {Chat{\'{e}}}, \citenamefont {Montagne},\ and\
  \citenamefont {qing Shi}}]{Ventejou2021}%
  \BibitemOpen
  \bibfield  {author} {\bibinfo {author} {\bibfnamefont {B.}~\bibnamefont
  {Ventejou}}, \bibinfo {author} {\bibfnamefont {H.}~\bibnamefont
  {Chat{\'{e}}}}, \bibinfo {author} {\bibfnamefont {R.}~\bibnamefont
  {Montagne}},\ and\ \bibinfo {author} {\bibfnamefont {X.}~\bibnamefont {qing
  Shi}},\ }\bibfield  {title} {\bibinfo {title} {Susceptibility of
  orientationally ordered active matter to chirality disorder},\ }\bibfield
  {journal} {\bibinfo  {journal} {Phys. Rev. Lett.}\ }\textbf {\bibinfo
  {volume} {127}},\ \href {https://doi.org/10.1103/physrevlett.127.238001}
  {10.1103/physrevlett.127.238001} (\bibinfo {year} {2021})\BibitemShut
  {NoStop}%
\bibitem [{\citenamefont {Kruk}\ \emph {et~al.}(2020)\citenamefont {Kruk},
  \citenamefont {Carrillo},\ and\ \citenamefont {Koeppl}}]{Kruk2020}%
  \BibitemOpen
  \bibfield  {author} {\bibinfo {author} {\bibfnamefont {N.}~\bibnamefont
  {Kruk}}, \bibinfo {author} {\bibfnamefont {J.~A.}\ \bibnamefont {Carrillo}},\
  and\ \bibinfo {author} {\bibfnamefont {H.}~\bibnamefont {Koeppl}},\
  }\bibfield  {title} {\bibinfo {title} {Traveling bands, clouds, and vortices
  of chiral active matter},\ }\bibfield  {journal} {\bibinfo  {journal} {Phys.
  Rev. E}\ }\textbf {\bibinfo {volume} {102}},\ \href
  {https://doi.org/10.1103/physreve.102.022604} {10.1103/physreve.102.022604}
  (\bibinfo {year} {2020})\BibitemShut {NoStop}%
\bibitem [{\citenamefont {Deblais}\ \emph {et~al.}(2018)\citenamefont
  {Deblais}, \citenamefont {Barois}, \citenamefont {Guerin}, \citenamefont
  {Delville}, \citenamefont {Vaudaine}, \citenamefont {Lintuvuori},
  \citenamefont {Boudet}, \citenamefont {Baret},\ and\ \citenamefont
  {Kellay}}]{Deblais2018}%
  \BibitemOpen
  \bibfield  {author} {\bibinfo {author} {\bibfnamefont {A.}~\bibnamefont
  {Deblais}}, \bibinfo {author} {\bibfnamefont {T.}~\bibnamefont {Barois}},
  \bibinfo {author} {\bibfnamefont {T.}~\bibnamefont {Guerin}}, \bibinfo
  {author} {\bibfnamefont {P.}~\bibnamefont {Delville}}, \bibinfo {author}
  {\bibfnamefont {R.}~\bibnamefont {Vaudaine}}, \bibinfo {author}
  {\bibfnamefont {J.}~\bibnamefont {Lintuvuori}}, \bibinfo {author}
  {\bibfnamefont {J.}~\bibnamefont {Boudet}}, \bibinfo {author} {\bibfnamefont
  {J.}~\bibnamefont {Baret}},\ and\ \bibinfo {author} {\bibfnamefont
  {H.}~\bibnamefont {Kellay}},\ }\bibfield  {title} {\bibinfo {title}
  {Boundaries control collective dynamics of inertial self-propelled robots},\
  }\bibfield  {journal} {\bibinfo  {journal} {Phys. Rev. Lett.}\ }\textbf
  {\bibinfo {volume} {120}},\ \href
  {https://doi.org/10.1103/physrevlett.120.188002}
  {10.1103/physrevlett.120.188002} (\bibinfo {year} {2018})\BibitemShut
  {NoStop}%
\bibitem [{\citenamefont {Vicsek}\ \emph {et~al.}(1995)\citenamefont {Vicsek},
  \citenamefont {Czir\'ok}, \citenamefont {Ben-Jacob}, \citenamefont {Cohen},\
  and\ \citenamefont {Shochet}}]{PhysRevLett.75.1226}%
  \BibitemOpen
  \bibfield  {author} {\bibinfo {author} {\bibfnamefont {T.}~\bibnamefont
  {Vicsek}}, \bibinfo {author} {\bibfnamefont {A.}~\bibnamefont {Czir\'ok}},
  \bibinfo {author} {\bibfnamefont {E.}~\bibnamefont {Ben-Jacob}}, \bibinfo
  {author} {\bibfnamefont {I.}~\bibnamefont {Cohen}},\ and\ \bibinfo {author}
  {\bibfnamefont {O.}~\bibnamefont {Shochet}},\ }\bibfield  {title} {\bibinfo
  {title} {Novel type of phase transition in a system of self-driven
  particles},\ }\href {https://doi.org/10.1103/PhysRevLett.75.1226} {\bibfield
  {journal} {\bibinfo  {journal} {Phys. Rev. Lett.}\ }\textbf {\bibinfo
  {volume} {75}},\ \bibinfo {pages} {1226} (\bibinfo {year}
  {1995})}\BibitemShut {NoStop}%
\bibitem [{\citenamefont {Wensink}\ and\ \citenamefont
  {L\"{o}wen}(lack)}]{PhysRevE.78.031409}%
  \BibitemOpen
  \bibfield  {author} {\bibinfo {author} {\bibfnamefont {H.~H.}\ \bibnamefont
  {Wensink}}\ and\ \bibinfo {author} {\bibfnamefont {H.}~\bibnamefont
  {L\"{o}wen}},\ }\bibfield  {title} {\bibinfo {title} {Aggregation of
  self-propelled colloidal rods near confining walls},\ }\href
  {https://doi.org/10.1103/PhysRevE.78.031409} {\bibfield  {journal} {\bibinfo
  {journal} {Phys. Rev. E}\ }\textbf {\bibinfo {volume} {78}},\ \bibinfo
  {pages} {031409} (\bibinfo {year} {2008\color{black}})}\BibitemShut {NoStop}%
\bibitem [{\citenamefont {Elgeti}\ and\ \citenamefont
  {Gompper}(lack)}]{Elgeti_2013}%
  \BibitemOpen
  \bibfield  {author} {\bibinfo {author} {\bibfnamefont {J.}~\bibnamefont
  {Elgeti}}\ and\ \bibinfo {author} {\bibfnamefont {G.}~\bibnamefont
  {Gompper}},\ }\bibfield  {title} {\bibinfo {title} {Wall accumulation of
  self-propelled spheres},\ }\href
  {https://doi.org/10.1209/0295-5075/101/48003} {\bibfield  {journal} {\bibinfo
   {journal} {Europhysics Letters}\ }\textbf {\bibinfo {volume} {101}},\
  \bibinfo {pages} {48003} (\bibinfo {year} {2013\color{black}})}\BibitemShut
  {NoStop}%
\bibitem [{sup()}]{supp}%
  \BibitemOpen
  \bibinfo {title} {See supplemental material at for simulations of unconfined
  chiral active matter systems, chiral confined active matter systems with
  different initial conditions, other order parameters and miscellaneous
  quantitative analysis}\BibitemShut {NoStop}%
\bibitem [{\citenamefont {Wioland}\ \emph
  {et~al.}(2013{\natexlab{b}})\citenamefont {Wioland}, \citenamefont
  {Woodhouse}, \citenamefont {Dunkel}, \citenamefont {Kessler},\ and\
  \citenamefont {Goldstein}}]{PhysRevLett.110.268102}%
  \BibitemOpen
\bibfield  {title} {  }\bibfield  {author} {\bibinfo {author} {\bibfnamefont
  {H.}~\bibnamefont {Wioland}}, \bibinfo {author} {\bibfnamefont {F.~G.}\
  \bibnamefont {Woodhouse}}, \bibinfo {author} {\bibfnamefont {J.}~\bibnamefont
  {Dunkel}}, \bibinfo {author} {\bibfnamefont {J.~O.}\ \bibnamefont
  {Kessler}},\ and\ \bibinfo {author} {\bibfnamefont {R.~E.}\ \bibnamefont
  {Goldstein}},\ }\bibfield  {title} {\bibinfo {title} {Confinement stabilizes
  a bacterial suspension into a spiral vortex},\ }\href
  {https://doi.org/10.1103/PhysRevLett.110.268102} {\bibfield  {journal}
  {\bibinfo  {journal} {Phys. Rev. Lett.}\ }\textbf {\bibinfo {volume} {110}},\
  \bibinfo {pages} {268102} (\bibinfo {year} {2013}{\natexlab{b}})}\BibitemShut
  {NoStop}%
\bibitem [{\citenamefont {Lei}\ \emph {et~al.}(lack)\citenamefont {Lei},
  \citenamefont {Zhao}, \citenamefont {Yan},\ and\ \citenamefont
  {Zhao}}]{D2SM01402E}%
  \BibitemOpen
  \bibfield  {author} {\bibinfo {author} {\bibfnamefont {T.}~\bibnamefont
  {Lei}}, \bibinfo {author} {\bibfnamefont {C.}~\bibnamefont {Zhao}}, \bibinfo
  {author} {\bibfnamefont {R.}~\bibnamefont {Yan}},\ and\ \bibinfo {author}
  {\bibfnamefont {N.}~\bibnamefont {Zhao}},\ }\bibfield  {title} {\bibinfo
  {title} {Collective behavior of chiral active particles with anisotropic
  interactions in a confined space},\ }\href
  {https://doi.org/10.1039/d2sm01402e} {\bibfield  {journal} {\bibinfo
  {journal} {Soft Matter}\ }\textbf {\bibinfo {volume} {19}},\ \bibinfo {pages}
  {1312} (\bibinfo {year} {2023\color{black}})}\BibitemShut {NoStop}%
\bibitem [{\citenamefont {Banerjee}\ \emph {et~al.}(2017)\citenamefont
  {Banerjee}, \citenamefont {Souslov}, \citenamefont {Abanov},\ and\
  \citenamefont {Vitelli}}]{vitelli2017}%
  \BibitemOpen
  \bibfield  {author} {\bibinfo {author} {\bibfnamefont {D.}~\bibnamefont
  {Banerjee}}, \bibinfo {author} {\bibfnamefont {A.}~\bibnamefont {Souslov}},
  \bibinfo {author} {\bibfnamefont {A.~G.}\ \bibnamefont {Abanov}},\ and\
  \bibinfo {author} {\bibfnamefont {V.}~\bibnamefont {Vitelli}},\ }\bibfield
  {title} {\bibinfo {title} {Odd viscosity in chiral active fluids},\
  }\bibfield  {journal} {\bibinfo  {journal} {Nat. Commun.}\ }\textbf {\bibinfo
  {volume} {8}},\ \href {https://doi.org/10.1038/s41467-017-01378-7}
  {10.1038/s41467-017-01378-7} (\bibinfo {year} {2017})\BibitemShut {NoStop}%
\bibitem [{\citenamefont {Lou}\ \emph {et~al.}(2022)\citenamefont {Lou},
  \citenamefont {Yang}, \citenamefont {Ding}, \citenamefont {Liu},
  \citenamefont {Chen}, \citenamefont {Zhou}, \citenamefont {Ye}, \citenamefont
  {Podgornik},\ and\ \citenamefont {Yang}}]{yan2022}%
  \BibitemOpen
  \bibfield  {author} {\bibinfo {author} {\bibfnamefont {X.}~\bibnamefont
  {Lou}}, \bibinfo {author} {\bibfnamefont {Q.}~\bibnamefont {Yang}}, \bibinfo
  {author} {\bibfnamefont {Y.}~\bibnamefont {Ding}}, \bibinfo {author}
  {\bibfnamefont {P.}~\bibnamefont {Liu}}, \bibinfo {author} {\bibfnamefont
  {K.}~\bibnamefont {Chen}}, \bibinfo {author} {\bibfnamefont {X.}~\bibnamefont
  {Zhou}}, \bibinfo {author} {\bibfnamefont {F.}~\bibnamefont {Ye}}, \bibinfo
  {author} {\bibfnamefont {R.}~\bibnamefont {Podgornik}},\ and\ \bibinfo
  {author} {\bibfnamefont {M.}~\bibnamefont {Yang}},\ }\bibfield  {title}
  {\bibinfo {title} {Odd viscosity-induced hall-like transport of an active
  chiral fluid},\ }\href {https://doi.org/10.1073/pnas.2201279119} {\bibfield
  {journal} {\bibinfo  {journal} {Proc. Natl. Acad. Sci. U.S.A.}\ }\textbf
  {\bibinfo {volume} {119}},\ \bibinfo {pages} {e2201279119} (\bibinfo {year}
  {2022})}\BibitemShut {NoStop}%
\bibitem [{\citenamefont {Hosaka}\ \emph {et~al.}(2021)\citenamefont {Hosaka},
  \citenamefont {Komura},\ and\ \citenamefont {Andelman}}]{komura2021}%
  \BibitemOpen
  \bibfield  {author} {\bibinfo {author} {\bibfnamefont {Y.}~\bibnamefont
  {Hosaka}}, \bibinfo {author} {\bibfnamefont {S.}~\bibnamefont {Komura}},\
  and\ \bibinfo {author} {\bibfnamefont {D.}~\bibnamefont {Andelman}},\
  }\bibfield  {title} {\bibinfo {title} {Hydrodynamic lift of a two-dimensional
  liquid domain with odd viscosity},\ }\href
  {https://doi.org/10.1103/PhysRevE.104.064613} {\bibfield  {journal} {\bibinfo
   {journal} {Phys. Rev. E}\ }\textbf {\bibinfo {volume} {104}},\ \bibinfo
  {pages} {064613} (\bibinfo {year} {2021})}\BibitemShut {NoStop}%
\bibitem [{\citenamefont {Shamipour}\ \emph {et~al.}(2021)\citenamefont
  {Shamipour}, \citenamefont {Caballero-Mancebo},\ and\ \citenamefont
  {Heisenberg}}]{heisenberg2021}%
  \BibitemOpen
  \bibfield  {author} {\bibinfo {author} {\bibfnamefont {S.}~\bibnamefont
  {Shamipour}}, \bibinfo {author} {\bibfnamefont {S.}~\bibnamefont
  {Caballero-Mancebo}},\ and\ \bibinfo {author} {\bibfnamefont {C.-P.}\
  \bibnamefont {Heisenberg}},\ }\bibfield  {title} {\bibinfo {title}
  {Cytoplasm's got moves},\ }\href
  {https://doi.org/10.1016/j.devcel.2020.12.002} {\bibfield  {journal}
  {\bibinfo  {journal} {Dev. Cell}\ }\textbf {\bibinfo {volume} {56}},\
  \bibinfo {pages} {213} (\bibinfo {year} {2021})}\BibitemShut {NoStop}%
\bibitem [{\citenamefont {Tee}\ \emph {et~al.}(2015)\citenamefont {Tee},
  \citenamefont {Shemesh}, \citenamefont {Thiagarajan}, \citenamefont
  {Hariadi}, \citenamefont {Anderson}, \citenamefont {Page}, \citenamefont
  {Volkmann}, \citenamefont {Hanein}, \citenamefont {Sivaramakrishnan},
  \citenamefont {Kozlov},\ and\ \citenamefont {Bershadsky}}]{Tee2015}%
  \BibitemOpen
  \bibfield  {author} {\bibinfo {author} {\bibfnamefont {Y.~H.}\ \bibnamefont
  {Tee}}, \bibinfo {author} {\bibfnamefont {T.}~\bibnamefont {Shemesh}},
  \bibinfo {author} {\bibfnamefont {V.}~\bibnamefont {Thiagarajan}}, \bibinfo
  {author} {\bibfnamefont {R.~F.}\ \bibnamefont {Hariadi}}, \bibinfo {author}
  {\bibfnamefont {K.~L.}\ \bibnamefont {Anderson}}, \bibinfo {author}
  {\bibfnamefont {C.}~\bibnamefont {Page}}, \bibinfo {author} {\bibfnamefont
  {N.}~\bibnamefont {Volkmann}}, \bibinfo {author} {\bibfnamefont
  {D.}~\bibnamefont {Hanein}}, \bibinfo {author} {\bibfnamefont
  {S.}~\bibnamefont {Sivaramakrishnan}}, \bibinfo {author} {\bibfnamefont
  {M.~M.}\ \bibnamefont {Kozlov}},\ and\ \bibinfo {author} {\bibfnamefont
  {A.~D.}\ \bibnamefont {Bershadsky}},\ }\bibfield  {title} {\bibinfo {title}
  {Cellular chirality arising from the self-organization of the actin
  cytoskeleton},\ }\href {https://doi.org/10.1038/ncb3137} {\bibfield
  {journal} {\bibinfo  {journal} {Nat. Cell Biol.}\ }\textbf {\bibinfo {volume}
  {17}},\ \bibinfo {pages} {445} (\bibinfo {year} {2015})}\BibitemShut
  {NoStop}%
\bibitem [{\citenamefont {Wan}\ \emph {et~al.}(2011)\citenamefont {Wan},
  \citenamefont {Ronaldson}, \citenamefont {Park}, \citenamefont {Taylor},
  \citenamefont {Zhang}, \citenamefont {Gimble},\ and\ \citenamefont
  {Vunjak-Novakovic}}]{Wan2011}%
  \BibitemOpen
  \bibfield  {author} {\bibinfo {author} {\bibfnamefont {L.~Q.}\ \bibnamefont
  {Wan}}, \bibinfo {author} {\bibfnamefont {K.}~\bibnamefont {Ronaldson}},
  \bibinfo {author} {\bibfnamefont {M.}~\bibnamefont {Park}}, \bibinfo {author}
  {\bibfnamefont {G.}~\bibnamefont {Taylor}}, \bibinfo {author} {\bibfnamefont
  {Y.}~\bibnamefont {Zhang}}, \bibinfo {author} {\bibfnamefont {J.~M.}\
  \bibnamefont {Gimble}},\ and\ \bibinfo {author} {\bibfnamefont
  {G.}~\bibnamefont {Vunjak-Novakovic}},\ }\bibfield  {title} {\bibinfo {title}
  {Micropatterned mammalian cells exhibit phenotype-specific left-right
  asymmetry},\ }\href {https://doi.org/10.1073/pnas.1103834108} {\bibfield
  {journal} {\bibinfo  {journal} {Proc. Natl. Acad. Sci. U.S.A.}\ }\textbf
  {\bibinfo {volume} {108}},\ \bibinfo {pages} {12295} (\bibinfo {year}
  {2011})}\BibitemShut {NoStop}%
\bibitem [{\citenamefont {Yashunsky}\ \emph {et~al.}(2022)\citenamefont
  {Yashunsky}, \citenamefont {Pearce}, \citenamefont {Blanch-Mercader},
  \citenamefont {Ascione}, \citenamefont {Silberzan},\ and\ \citenamefont
  {Giomi}}]{giomi2022}%
  \BibitemOpen
  \bibfield  {author} {\bibinfo {author} {\bibfnamefont {V.}~\bibnamefont
  {Yashunsky}}, \bibinfo {author} {\bibfnamefont {D.~J.~G.}\ \bibnamefont
  {Pearce}}, \bibinfo {author} {\bibfnamefont {C.}~\bibnamefont
  {Blanch-Mercader}}, \bibinfo {author} {\bibfnamefont {F.}~\bibnamefont
  {Ascione}}, \bibinfo {author} {\bibfnamefont {P.}~\bibnamefont {Silberzan}},\
  and\ \bibinfo {author} {\bibfnamefont {L.}~\bibnamefont {Giomi}},\ }\bibfield
   {title} {\bibinfo {title} {Chiral edge current in nematic cell monolayers},\
  }\href {https://doi.org/10.1103/PhysRevX.12.041017} {\bibfield  {journal}
  {\bibinfo  {journal} {Phys. Rev. X}\ }\textbf {\bibinfo {volume} {12}},\
  \bibinfo {pages} {041017} (\bibinfo {year} {2022})}\BibitemShut {NoStop}%
\bibitem [{\citenamefont {Drescher}\ \emph {et~al.}(2009)\citenamefont
  {Drescher}, \citenamefont {Leptos}, \citenamefont {Tuval}, \citenamefont
  {Ishikawa}, \citenamefont {Pedley},\ and\ \citenamefont
  {Goldstein}}]{goldstein2009}%
  \BibitemOpen
  \bibfield  {author} {\bibinfo {author} {\bibfnamefont {K.}~\bibnamefont
  {Drescher}}, \bibinfo {author} {\bibfnamefont {K.~C.}\ \bibnamefont
  {Leptos}}, \bibinfo {author} {\bibfnamefont {I.}~\bibnamefont {Tuval}},
  \bibinfo {author} {\bibfnamefont {T.}~\bibnamefont {Ishikawa}}, \bibinfo
  {author} {\bibfnamefont {T.~J.}\ \bibnamefont {Pedley}},\ and\ \bibinfo
  {author} {\bibfnamefont {R.~E.}\ \bibnamefont {Goldstein}},\ }\bibfield
  {title} {\bibinfo {title} {Dancing volvox: Hydrodynamic bound states of
  swimming algae},\ }\href {https://doi.org/10.1103/PhysRevLett.102.168101}
  {\bibfield  {journal} {\bibinfo  {journal} {Phys. Rev. Lett.}\ }\textbf
  {\bibinfo {volume} {102}},\ \bibinfo {pages} {168101} (\bibinfo {year}
  {2009})}\BibitemShut {NoStop}%
\bibitem [{\citenamefont {Petroff}\ \emph {et~al.}(2015)\citenamefont
  {Petroff}, \citenamefont {Wu},\ and\ \citenamefont
  {Libchaber}}]{libchaber2015}%
  \BibitemOpen
  \bibfield  {author} {\bibinfo {author} {\bibfnamefont {A.~P.}\ \bibnamefont
  {Petroff}}, \bibinfo {author} {\bibfnamefont {X.-L.}\ \bibnamefont {Wu}},\
  and\ \bibinfo {author} {\bibfnamefont {A.}~\bibnamefont {Libchaber}},\
  }\bibfield  {title} {\bibinfo {title} {Fast-moving bacteria self-organize
  into active two-dimensional crystals of rotating cells},\ }\href
  {https://doi.org/10.1103/PhysRevLett.114.158102} {\bibfield  {journal}
  {\bibinfo  {journal} {Phys. Rev. Lett.}\ }\textbf {\bibinfo {volume} {114}},\
  \bibinfo {pages} {158102} (\bibinfo {year} {2015})}\BibitemShut {NoStop}%
\bibitem [{\citenamefont {Tan}\ \emph {et~al.}(2022)\citenamefont {Tan},
  \citenamefont {Mietke}, \citenamefont {Li}, \citenamefont {Chen},
  \citenamefont {Higinbotham}, \citenamefont {Foster}, \citenamefont {Gokhale},
  \citenamefont {Dunkel},\ and\ \citenamefont {Fakhri}}]{fakhri2022}%
  \BibitemOpen
  \bibfield  {author} {\bibinfo {author} {\bibfnamefont {T.~H.}\ \bibnamefont
  {Tan}}, \bibinfo {author} {\bibfnamefont {A.}~\bibnamefont {Mietke}},
  \bibinfo {author} {\bibfnamefont {J.}~\bibnamefont {Li}}, \bibinfo {author}
  {\bibfnamefont {Y.}~\bibnamefont {Chen}}, \bibinfo {author} {\bibfnamefont
  {H.}~\bibnamefont {Higinbotham}}, \bibinfo {author} {\bibfnamefont {P.~J.}\
  \bibnamefont {Foster}}, \bibinfo {author} {\bibfnamefont {S.}~\bibnamefont
  {Gokhale}}, \bibinfo {author} {\bibfnamefont {J.}~\bibnamefont {Dunkel}},\
  and\ \bibinfo {author} {\bibfnamefont {N.}~\bibnamefont {Fakhri}},\
  }\bibfield  {title} {\bibinfo {title} {Odd dynamics of living chiral
  crystals},\ }\href {https://doi.org/10.1038/s41586-022-04889-6} {\bibfield
  {journal} {\bibinfo  {journal} {Nature}\ }\textbf {\bibinfo {volume} {607}},\
  \bibinfo {pages} {287} (\bibinfo {year} {2022})}\BibitemShut {NoStop}%
\bibitem [{\citenamefont {Harris}\ \emph {et~al.}(lack)\citenamefont {Harris},
  \citenamefont {Millman}, \citenamefont {van~der Walt}, \citenamefont
  {Gommers}, \citenamefont {Virtanen}, \citenamefont {Cournapeau},
  \citenamefont {Wieser}, \citenamefont {Taylor}, \citenamefont {Berg},
  \citenamefont {Smith}, \citenamefont {Kern}, \citenamefont {Picus},
  \citenamefont {Hoyer}, \citenamefont {van Kerkwijk}, \citenamefont {Brett},
  \citenamefont {Haldane}, \citenamefont {del R{\'{i}}o}, \citenamefont
  {Wiebe}, \citenamefont {Peterson}, \citenamefont {G{\'{e}}rard-Marchant},
  \citenamefont {Sheppard}, \citenamefont {Reddy}, \citenamefont {Weckesser},
  \citenamefont {Abbasi}, \citenamefont {Gohlke},\ and\ \citenamefont
  {Oliphant}}]{harris2020array}%
  \BibitemOpen
  \bibfield  {author} {\bibinfo {author} {\bibfnamefont {C.~R.}\ \bibnamefont
  {Harris}}, \bibinfo {author} {\bibfnamefont {K.~J.}\ \bibnamefont {Millman}},
  \bibinfo {author} {\bibfnamefont {S.~J.}\ \bibnamefont {van~der Walt}},
  \bibinfo {author} {\bibfnamefont {R.}~\bibnamefont {Gommers}}, \bibinfo
  {author} {\bibfnamefont {P.}~\bibnamefont {Virtanen}}, \bibinfo {author}
  {\bibfnamefont {D.}~\bibnamefont {Cournapeau}}, \bibinfo {author}
  {\bibfnamefont {E.}~\bibnamefont {Wieser}}, \bibinfo {author} {\bibfnamefont
  {J.}~\bibnamefont {Taylor}}, \bibinfo {author} {\bibfnamefont
  {S.}~\bibnamefont {Berg}}, \bibinfo {author} {\bibfnamefont {N.~J.}\
  \bibnamefont {Smith}}, \bibinfo {author} {\bibfnamefont {R.}~\bibnamefont
  {Kern}}, \bibinfo {author} {\bibfnamefont {M.}~\bibnamefont {Picus}},
  \bibinfo {author} {\bibfnamefont {S.}~\bibnamefont {Hoyer}}, \bibinfo
  {author} {\bibfnamefont {M.~H.}\ \bibnamefont {van Kerkwijk}}, \bibinfo
  {author} {\bibfnamefont {M.}~\bibnamefont {Brett}}, \bibinfo {author}
  {\bibfnamefont {A.}~\bibnamefont {Haldane}}, \bibinfo {author} {\bibfnamefont
  {J.~F.}\ \bibnamefont {del R{\'{i}}o}}, \bibinfo {author} {\bibfnamefont
  {M.}~\bibnamefont {Wiebe}}, \bibinfo {author} {\bibfnamefont
  {P.}~\bibnamefont {Peterson}}, \bibinfo {author} {\bibfnamefont
  {P.}~\bibnamefont {G{\'{e}}rard-Marchant}}, \bibinfo {author} {\bibfnamefont
  {K.}~\bibnamefont {Sheppard}}, \bibinfo {author} {\bibfnamefont
  {T.}~\bibnamefont {Reddy}}, \bibinfo {author} {\bibfnamefont
  {W.}~\bibnamefont {Weckesser}}, \bibinfo {author} {\bibfnamefont
  {H.}~\bibnamefont {Abbasi}}, \bibinfo {author} {\bibfnamefont
  {C.}~\bibnamefont {Gohlke}},\ and\ \bibinfo {author} {\bibfnamefont {T.~E.}\
  \bibnamefont {Oliphant}},\ }\bibfield  {title} {\bibinfo {title} {Array
  programming with {NumPy}},\ }\href
  {https://doi.org/10.1038/s41586-020-2649-2} {\bibfield  {journal} {\bibinfo
  {journal} {Nature}\ }\textbf {\bibinfo {volume} {585}},\ \bibinfo {pages}
  {357} (\bibinfo {year} {2020\color{black}})}\BibitemShut {NoStop}%
\end{thebibliography}%

\end{document}